\documentclass[reprint, superscriptaddress, amsmath, amssymb,nofootinbib, aps, prl]{revtex4-1}
\usepackage[dvipdfmx]{graphicx}
\usepackage{dcolumn}
\usepackage{bm}
\usepackage[dvipdfmx]{hyperref}
\usepackage{comment}
\usepackage{physics}
\usepackage{braket}

\begin{document}
\preprint{APS/123-QED}
\title{Oscillating-charged Andreev Bound States and Their Appearance in UTe$_2$}
\author{Satoshi Ando}
\thanks{These authors contributed equally to this work.}
\affiliation{Department of Applied Physics, Nagoya University, Nagoya 464-8603, Japan}
\author{Shingo Kobayashi}
\thanks{These authors contributed equally to this work.}
\affiliation{RIKEN Center for Emergent Matter Science, Wako, Saitama, 351-0198, Japan}
\author{Andreas P. Schnyder}
\affiliation{Max-Planck-Institut f\"ur Festk\"orperforschung, Heisenbergstrasse 1, D-70569 Stuttgart, Germany}
\author{Yasuhiro Asano}
\affiliation{Department of Applied Physics, Hokkaido University, Sapporo 060-8628, Japan}
\author{Satoshi Ikegaya}
\thanks{These authors contributed equally to this work.}
\affiliation{Department of Applied Physics, Nagoya University, Nagoya 464-8603, Japan}
\affiliation{Institute for Advanced Research, Nagoya University, Nagoya 464-8601, Japan}
\date{\today}
\begin{abstract}
Surface Andreev bound states, including Majorana bound states in topological superconductors, are typically charge neutral.
In this work, we demonstrate the emergence of unconventional charged Andreev bound states in a superconductor with a sublattice degree of freedom,
where the sign of charge density of the Andreev bound states oscillates between the two sublattices.
The oscillating-charged Andreev bound states lead to a complete breakdown of the proportionality
among the electron part of the spectral function, the local density of states, and the tunneling conductance spectrum for energies below the superconducting gap.
We also discuss the possible occurrence of these Andreev bound states in UTe$_2$ and locally noncentrosymmetric superconductors.
\end{abstract}
\maketitle

\textit{Introduction}.
Andreev bound states (ABSs) are distinctive subgap states that arise from the quantum coherence between electrons and holes in superconductors~\cite{andreev_64,andreev_65}.
To date, ABSs have played a crucial role in understanding and advancing the physics of superconductivity.
For example, the supercurrent in Josephson junctions is governed by the ABSs formed at the junction interface~\cite{beenakker_91,tanaka_00(r),golubov_04,kwon_04}.
Superconductors with anisotropic pairing inherently host gapless/mid-gap ABSs at their surfaces~\cite{buchholtz_81,hu_94,kashiwaya_96,asano_04}.
The zero-bias conductance peak induced by these \emph{surface} ABSs is widely recognized as the evidence for unconventional superconductivity%
~\cite{bruder_90,tanaka_95,tanaka_96,tanaka_00,sengupta_01,sengupta_02}.
Currently, Majorana bound states in topological superconductors~\cite{green_00,kitaev_01,wilczek_09,kane_10,zhang_11,schnyder_16,flensberg_12,nagaosa_12,sato_16,sato_17}%
---a special type of ABSs composed of quasiparticles that are their own antiparticles---%
have attracted considerable attention, due to their potential applications in fault-tolerant topological quantum computing%
~\cite{ivanov_01,kitaev_03,sarma_08,alicea_16,fisher_11,tewari_11,sarma_15}.

In this Letter, we propose a novel class of ABSs that exhibit a unique charge property.
Note that the vast majority of surface ABSs studied so far, including Majorana bound states, are charge neutral.
In contrast, we show that a superconductor with a sublattice degree of freedom can host unconventional surface ABSs whose charge density oscillates in sign between the two sublattices.
Because of this property, we refer to these ABSs as oscillating-charged ABSs (OCABSs).
It is well known that, in the presence of charge-neutral ABSs,
the electron part of the spectral function $\rho_e$, the local density of states $\rho$, and the tunneling conductance spectrum $G$
are nearly proportional to one another for energies below the superconducting gap.
However, we show that the OCABSs lead to a complete breaking of this well-known proportionality.
In addition, we describe the possible appearance of OCABSs in the superconducting UTe$_2$, which has recently received immense attention due to its field-induced reentrant spin-triplet superconductivity%
~\cite{ran_19,metz_19,sundar_19,aoki_19,madhavan_20,hayes_21,duan_21,aoki_22,agterberg_21}.
We also discuss a significant relationship between the OCABS and a quantum geometry arising from the inter-sublattice pairing.
\begin{figure}[h]
\begin{center}
\includegraphics[width=0.499\textwidth]{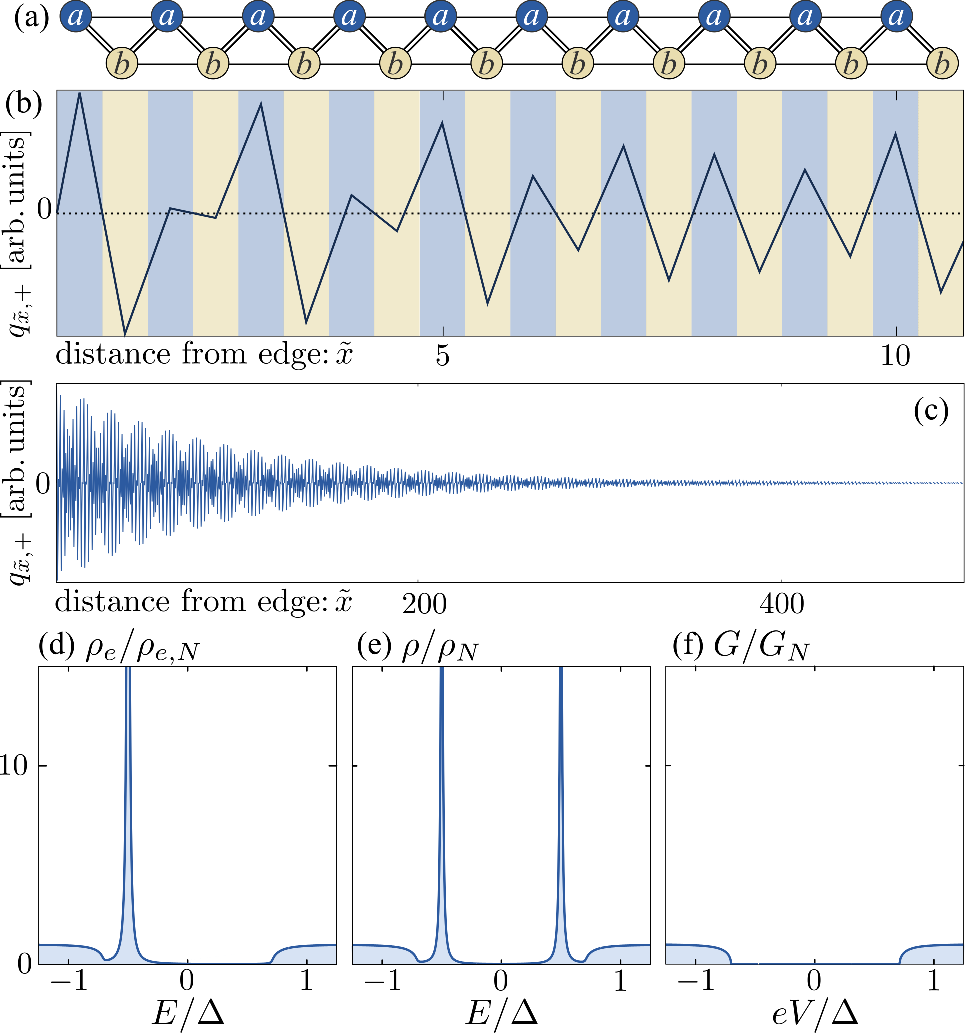}
\caption{\label{fig:1}%
(a) Schematic image of the sublattice superconductor. The single (double) line represents the coupling through $t$ ($\Delta$).
(b) and (c) Charge density of the positive energy OCABS, $q_{\tilde{x},+}$, as a function of the distance from the edge, with $\mu=0.1t$ and $\Delta=0.01t$.
In (b) and (c), the results for $1\leq \tilde{x} \leq 10+\frac{1}{2}$ and $1\leq \tilde{x} \leq 500+\frac{1}{2}$ are plotted, respectively.
For the positive energy OCABSs, the sign of the charge density on the $a$-sublattices ($b$-sublattices) is fixed to be positive (negative).
(d) $\rho_e(1,E)$  and (e) $\rho(1,E)$ as a function of the energy.
(f) $G(eV)$ as a function of the bias voltage.}
\end{center}
\end{figure}

\textit{Minimal model}.
We introduce a generic model with sublattice degrees of freedom that exhibits OCABSs.
Specifically, we consider a sublattice superconductor shown in Fig.~\ref{fig:1}(a).
There are two sites within the primitive unit cells, i.e., the $a$ and $b$ sites located at $x$ and $x+\frac{1}{2}a_0$, respectively, forming two distinct sublattices. 
In the following, we set the lattice constant $a_0$ to one, so that $x$ takes on integer values. 
For simplicity, we first consider the extreme case, where finite hopping occurs only between sites of the same sublattice
[single lines in Fig.~\ref{fig:1}(a)], while pairing occurs only between different sublattices [double lines in Fig.~\ref{fig:1}(a)].
This extreme limit may be unrealistic, but it  is useful to highlight the key properties of the OCABSs.
A straightforward extension of our minimal model describes a locally noncentrosymmetric superconductor~~\cite{sigrist_11,youn_14,yanase_23}
with an interlayer pairing~\cite{ramires_21(1),ramires_21(2)}, which is proposed to describe the superconductor CeRh$_2$As$_2$~\cite{hassinger_21,hassinger_22}.
In the Supplemental Material (SM)~\cite{sm}, we show the explicit BdG Hamiltonian for the locally noncentrosymmetric superconductor and demonstrate the appearance of the OCABSs.
Furthermore, we will later show that the low-energy excitations of the superconducting UTe$_2$~\cite{agterberg_21} are effectively described by our minimal model.
Let us consider the Bogoliubov-de Gennes (BdG) Hamiltonian in momentum space:
\begin{align}
\begin{split}
& H=\frac{1}{2}\sum_{k} \Psi^{\dagger}_{k} \check{H}_k \Psi_{k},\quad
\check{H}_k = \left[\begin{array}{cc} \hat{H}_{k,+}& 0 \\ 0 & \hat{H}_{k,-}\end{array}\right],\\
&\hat{H}_{k,s}=\left[\begin{array}{cc}t \cos (k) - \mu& \Delta e^{-is \frac{k}{2}} \cos (k/2) \\ \Delta e^{is \frac{k}{2}} \cos (k/2) & -t \cos (k) + \mu\end{array}\right], \\
&\Psi_{k}=[\psi_{k,a},\psi^{\dagger}_{-k,b}, \psi_{k,b},\psi^{\dagger}_{-k,a}]^{\mathrm{T}}
\label{eq:minimal_hm}
\end{split}
\end{align}
where $s=+1$ ($-1$),
$\psi_{k,\alpha}$ is the annihilation operator of an electron at sublattice $\alpha$ with momentum $k$, $t$ denotes the hopping integral, and $\mu$ represents the chemical potential.
The amplitude of the pair potential, which belongs to an inter-sublattice-even (spin-singlet) extended-$s$-wave pairing symmetry, is given by $\Delta$.
The particle-hole symmetry operator is given by
\begin{align}
\check{C}= \left[\begin{array}{cc}0 & \hat{C} \\ \hat{C} & 0\end{array}\right],\quad
\hat{C}= \left[\begin{array}{cc}0 & 1 \\ -1 & 0\end{array}\right] \mathcal{K},\quad
\end{align}
where $\mathcal{K}$ represents the complex conjugation operator and $\check{C}^2=-1$, due to the absence of spin degrees of freedom. 
The BdG Hamiltonian $\check{H}_k$ belongs to class CI in the Altland-Zirnbauer symmetry classes~\cite{altland_97}, which has no topological invariant in one dimension. 
The particle-hole symmetry is broken within each block component,
$\hat{C} \hat{H}_{k,s} \hat{C}^{-1} = - \hat{H}_{-k,-s} \neq - \hat{H}_{-k,s}$, while the operation of $\check{C}$ additionally interchanges the two blocks.
Namely, the block components $\hat{H}_{\boldsymbol{k},\pm}$ belongs to class AI (a symmetry class of spinless insulators).
We also note that the block component $\hat{H}_{k,s}$ has a form similar to the Hamiltonian of the Rice--Mele model~\cite{mele_82},
where the inter-sublattice hopping terms and the on-site potentials of the Rice--Mele model are replaced by the inter-sublattice pair potential and the normal-state kinetic term, respectively.
The Rice--Mele model is a fundamental model to describe dielectric materials that exhibit a bulk polarization.
At the surface, the polarization gives rise to an accumulation of probability density, which is primarily attributed to subgap states appearing at the surface%
~\cite{king-smith_93,vanderbilt_93,resta_94,resta_07,bardarson_17,ortix_17,watanabe_18,rhim_20,pletyukhov_20}.
Therefore, based on the similarity between the Rice-Mele model and our model, we expect the appearance of surface ABSs that contribute to the surface probability density.
To verify this expectation, we solve the BdG equation in real space:
\begin{align}
\hat{T}_{s}\varphi_{x-1,s}+\hat{T}^{\dagger}_{s}\varphi_{x+1,s}+\hat{K}\varphi_{x,s} = E_{s}\varphi_{x,s},
\end{align}
with
\begin{align}
\begin{split}
&\hat{T}_{s}=\left[\begin{array}{cc}\frac{t}{2} & \frac{(1+s)\Delta}{4} \\ \frac{(1-s)\Delta}{4} & -\frac{t}{2}\end{array}\right],\quad
\hat{K}=\left[\begin{array}{cc}-\mu & \frac{\Delta}{2} \\ \frac{\Delta}{2} & \mu\end{array}\right],\\
&\varphi_{x,+}=[u_{x,+},v_{x+\frac{1}{2},+}]^{\mathrm{T}}, \; \varphi_{x,-}=[u_{x+\frac{1}{2},-},v_{x,-}]^{\mathrm{T}},\\
&u_{x+\frac{1}{2},+}=v_{x,+}=u_{x,-}=v_{x+\frac{1}{2},-}=0,
\end{split}
\end{align}
where $u_{x,s}$ ($v_{x,s}$) denotes the electron (hole) component in the sublattice $a$,
and $u_{x+\frac{1}{2},s}$ ($v_{x+\frac{1}{2},s}$) represents the electron (hole) component in the sublattice $b$.
We consider the semi-infinite system for $x>0$ and apply an open boundary condition, $\varphi_{0,s}=0$.
As a result, we find the solutions of the ABSs for $|\mu|<t$, where the energy eigenvalue and eigenvector are given by
\begin{align}
E_s=-s \frac{\Delta}{2t}(t+\mu), \quad
\varphi_{x,s}=\left[\begin{array}{cc}1\\-s\end{array}\right]\phi_x,
\label{eq:ew_ocabs}
\end{align}
respectively.
The derivation of ABSs and the explicit form of the real function $\phi_x$ are provided in the SM~\cite{sm}.
Here, we focus on the charge density of the ABSs, which is given by
\begin{align}
q_{\tilde{x},s}=e(|u_{\tilde{x},s}|^2-|v_{\tilde{x},s}|^2),
\label{eq:cd}
\end{align}
with $\tilde{x}=x$ or $x+\frac{1}{2}$.
For ABSs of $\varphi_{x,+(-)}$, the electron component is present only at the $a$-sublattices ($b$-sublattices), while the hole component is present only at the $b$-sublattices ($a$-sublattices).
Therefore, we obtain
\begin{align}
\begin{split}
&q_{x,s}=e s|\phi_x|^2,\\
&q_{x+\frac{1}{2},s}=-e s|\phi_x|^2.
\end{split}
\end{align}
Remarkably, the charge density of the ABSs changes sign for each sublattice, as also shown in Figs.~\ref{fig:1}(b) and \ref{fig:1}(c).
This establishes the oscillating charge nature of the OCABS.

To highlight the intriguing charge characteristics of the OCABSs, we examine three quantities:
the electron part of the spectral function $\rho_e$, the local density of states $\rho$, and the tunneling conductance $G$.
Proportionality in these quantities is widely recognized as a general feature in the presence of charge neutral ABSs,
while most of surface ABSs studied to date, including Majorana bound states, are charge neutral.
However, we will show that the OCABSs break this proportionality completely.
We first examine $\rho_e$ and $\rho$, which are given by
\begin{align}
&\rho_e(\tilde{x},E)=-\frac{1}{\pi}\mathrm{Im}[g(\tilde{x},\tilde{x},E+i\delta)], \\
&\rho(\tilde{x},E)=-\frac{1}{\pi}\mathrm{Im}[g(\tilde{x},\tilde{x},E+i\delta)+\underline{g}(\tilde{x},\tilde{x},E+i\delta)],
\end{align}
where
\begin{align}
\mathcal{G}(\tilde{x},\tilde{x}^{\prime},E)=\left[ \begin{array}{cc}
g(\tilde{x},\tilde{x}^{\prime},E)&f(\tilde{x},\tilde{x}^{\prime},E) \\
\underline{f}(\tilde{x},\tilde{x}^{\prime},E)&\underline{g}(\tilde{x},\tilde{x}^{\prime},E)\end{array} \right],
\end{align}
represents the Gorkov--Green's function, and $\delta$ characterizes the broadening of the energy.
Here we assume that an ABS exists at an energy $E_B$, which is energetically well separated from other states,
and the wave function of the ABS is represented by $\varphi^B_{\tilde{x}}=[u^B_{\tilde{x}},v^T _{\tilde{x}}]^{\mathrm{T}}$.
By focusing on $E \approx E_B$ and extracting the contributions from the ABSs to the Green's function, we approximately obtain~\cite{sm}
\begin{align}
\begin{split}
& \rho_e(\tilde{x},E) \approx \frac{1}{\pi}\frac{\delta}{(E-E_B)^2 + \delta^2} n^e_{\tilde{x}},\\
& \rho(\tilde{x},E) \approx \frac{1}{\pi}\frac{\delta}{(E-E_B)^2 + \delta^2} n_{\tilde{x}},
\label{eq:rho_abs}
\end{split}
\end{align}
where $n_{\tilde{x}}=n^e_{\tilde{x}}+n^h_{\tilde{x}}$ with $n^e_{\tilde{x}}=|u^B_{\tilde{x}}|^2$ and $n^h_{\tilde{x}}=|v^B_{\tilde{x}}|^2$.
If the ABS is charge-neutral, i.e., $n^e_{\tilde{x}} = n^h_{\tilde{x}} = n_{\tilde{x}}/2$, the proportionality of $\rho(\tilde{x},E) = 2 \rho_e(\tilde{x},E)$ is satisfied.
However, for the OCABS with the negative (positive) energy, $n^e_{\tilde{x}}$ ($n^h_{\tilde{x}}$) is finite only in the $a$-sublattices ($b$-sublattices),
while $n^h_{\tilde{x}}$ ($n^e_{\tilde{x}}$) is finite only in the $b$-sublattices ($a$-sublattices).
As a result, at the outermost site belonging to the $a$-sublattice (i.e., $x=1$), $\rho_e(1,E) $ is finite only for negative energies, while $\rho(1,E)$ is finite for both positive and negative energies.
In Fig.~\ref{fig:1}(d) and Fig.~\ref{fig:1}(e), we show the numerically computed $\rho_e(1,E)$ and $\rho(1,E)$ as a function of energy.
For these calculations, we set $\mu=0$, $\Delta=0.001t$ and $\delta=10^{-2}\Delta$, and the results are normalized to the value calculated with $\Delta=E=0$.
The Green's functions are calculated using recursive Green's function techniques~\cite{fisher_81,ando_91}.
As expected, $\rho_e(1,E)$ has the asymmetric spectrum~\cite{fu_12,yada_22,fukaya_23} with the peak at $E=-|E_s|$, while $\rho(1,E)$ displays the symmetric peaks at $E=\pm |E_s|$.
We also remark that $\rho_e$ at the second outermost site (i.e., $x=1+\frac{1}{2}$) shows the peak only at $E=+|E_s|$, which manifests the oscillating charge nature of the OCABSs.
Next, we discuss the tunneling conductance $G$.
We consider a normal-metal lead attached to the outermost site of the superconductor (i.e., the $a$-sublattice at $x=1$).
The normal-metal lead is described by $H_N = -t\sum_x (a^{\dagger}_{x+1} a_x + \mathrm{h.c.})$,
where $a_x$ denotes the annihilation operator of an electron in the normal-metal, and $x$ are integer numbers for $x<0$.
The coupling between the lead and the superconductor is described by $H_T = -t_T (\psi^{\dagger}_{\tilde{x}=1} a_{x=0} + \mathrm{h.c.})$,
where $\psi_{\tilde{x}}$ denotes the annihilation operator of an electron at $\tilde{x}$ in the superconductor.
We calculate $G$ using the Blonder--Tinkham--Klapwijk formula~\cite{klapwijk_82},
\begin{align}
G (eV)= \frac{e^2}{h} [1-|b(E)|^2 +|a(E)|^2]_{E=eV},
\end{align}
where $a(E)$ and $b(E)$ are the Andreev and normal reflection coefficients at energy $E$, respectively.
In the presence of the ABS, the tunneling conductance at $eV \approx E_B$ is given by~\cite{beenakker_08,feigelman_13,flensberg_20}
\begin{align}
G(eV) \approx \frac{2e^2}{h}\frac{\Gamma^2}{(eV-E_B)^2 + \Gamma^2} \frac{4 n^e_1 n^h_1}{(n_1)^2},
\label{eq:gns}
\end{align}
where $\Gamma = (t_T^2/t) n_1$.
The detailed derivation of Eq.~(\ref{eq:gns}) is provided in the SM~\cite{sm}.
If the ABS is charge neutral (i.e., $n^e_{\tilde{x}} = n^h_{\tilde{x}} = n_{\tilde{x}}/2$), we obtain the well known proportionality of
\begin{align}
G(eV) = \frac{2e^2}{h} \frac{ \pi t_T^2}{t} \left. \rho(1,E) \right|_{E=eV}, \nonumber
\end{align}
where we set $\delta=\Gamma$.
For the OCABS, however, we expect $G=0$ because the electron and hole components do not exist at the same sites, i.e., $n^e_1 n^h_1=0$.
In Fig.~\ref{fig:1}(f), we show the numerically calculated $G$ as a function of the bias voltage.
We assume the low-transparency interface with $t_T=0.02t$, and the results are normalized to the normal conductance calculated by setting $\Delta=eV=0$.
The scattering coefficients are computed using recursive Green's function techniques~\cite{fisher_81,ando_91}.
As expected, $G$ does not exhibit any peaks for $|eV|<\Delta$.
Eventually, we confirm the distinct breakdown of the proportionality among $\rho_e$, $\rho$, and $G$ in the presence of the OCABSs.

In the minimal model, we implicitly assume a symmetry that enables us to decompose the BdG Hamiltonian into the block components $\hat{H}_{k,s}$.
This symmetry is broken by perturbations in the block-off-diagonal elements of $H_{\boldsymbol{k}}$, such as tunnel coupling between the $a$- and $b$-sublattices.
Nevertheless, as long as the perturbation is sufficiently weak, the OCABSs can still be expected to appear (while their energy levels are slightly shifted by the perturbation).
This tolerance of OCABSs allows us to expect their appearance in more realistic systems beyond the minimal model.
Indeed, in the following section, we discuss the possible appearance of OCABSs in the superconductor UTe$_2$.

\textit{Superconductor UTe$_2$}.
We consider the BdG Hamiltonian proposed in Ref.~[\onlinecite{agterberg_21}]:
\begin{align}
\begin{split}
&H = \frac{1}{2}\sum_{\boldsymbol{k},\sigma}\Psi^{\dagger}_{\boldsymbol{k}\sigma} \bar{H}_{\boldsymbol{k}} \Psi_{\boldsymbol{k}\sigma}, \quad
\bar{H}_{\boldsymbol{k}}=\left[\begin{array}{cc}\check{h}_{\boldsymbol{k}} & \check{h}_\Delta \\
\check{h}^{\dagger}_\Delta & -\check{h}^{\ast}_{-\boldsymbol{k}} \end{array}\right],\\
&\Psi^{\dagger}_{\boldsymbol{k}\sigma} = [\psi^{\dagger}_{\boldsymbol{k}\sigma},\psi^{\mathrm{T}}_{-\boldsymbol{k}\bar{\sigma}}],\\
&\psi^{\dagger}_{\boldsymbol{k}\sigma}=[\psi^{\dagger}_{\boldsymbol{k},\sigma,a,1},\psi^{\dagger}_{\boldsymbol{k},\sigma,a,2},
\psi^{\dagger}_{\boldsymbol{k},\sigma,b,1},\psi^{\dagger}_{\boldsymbol{k},\sigma,b,2}],\\
&\label{eq:ute2_hm1}
\end{split}
\end{align}
with
\begin{align}
\begin{split}
&\check{h}_{\boldsymbol{k}}=
\left[\begin{array}{cc} \hat{K}_{\boldsymbol{k}} & \hat{T}^{\dagger}_{\boldsymbol{k}} \\
\hat{T}_{\boldsymbol{k}}& \hat{K}_{\boldsymbol{k}} \end{array} \right],\quad
\hat{K}_{\boldsymbol{k}}=\left[ \begin{array}{cc} \epsilon_{\boldsymbol{k}} & m_g \\ m_g & \epsilon_{\boldsymbol{k}} \end{array}\right],\\
&\hat{T}_{\boldsymbol{k}}=e^{i\frac{k_x}{2}}e^{i\frac{k_y}{2}}e^{i\frac{k_z}{2}}
\left[ \begin{array}{cc} 0 & f_{g,\boldsymbol{k}}-if_{z,\boldsymbol{k}} \\ f_{g,\boldsymbol{k}}+if_{z,\boldsymbol{k}} & 0 \end{array}\right],\\
&\epsilon_{\boldsymbol{k}}=t_1 \cos (k_x) + t_2 \cos (k_y) -\mu,\\
&f_{g,\boldsymbol{k}}=t_3 \cos(k_x/2)\cos(k_y/2)\cos(k_z/2),\\
&f_{z,\boldsymbol{k}}=t_z \cos(k_x/2)\cos(k_y/2)\sin(k_z/2),\\
&\check{h}_{\Delta}=\left[\begin{array}{cc} \hat{\Delta} & 0 \\ 0 & \hat{\Delta} \end{array} \right], \quad
\hat{\Delta} = \left[ \begin{array}{cc} 0 & \Delta \\ -\Delta & 0 \end{array} \right],
\end{split}
\end{align}
where $\psi_{\boldsymbol{k},\sigma,\alpha,l}$ is the annihilation operator of an electron with momentum $\boldsymbol{k}$ with spin $\sigma$ at the $l$-th rung of the sublattice $\alpha$.
The index $\bar{\sigma}$ means the opposite spin of $\sigma$.
The rung degree of freedom is originated from the two inequivalent U sites that constitute the ladder rung.
Note that Eq.~(\ref{eq:ute2_hm1}) is modified to take into account the sublattice degrees of freedom explicitly.
The parameters are chosen as
$(\mu,$ $t_1,$ $t_2,$ $m_0,$ $t_3,$ $t_z)$$=$$(-0.129,$ $-0.0892,$ $0.0678,$ $-0.062,$ $0.0742,$ $-0.0742)$~\cite{agterberg_21}.
We assume the inter-rung-odd spin-triplet $s$-wave pair potential belonging to $A_u$ of the $D_{2h}$ point-group symmetry, where we choose $\Delta=0.001$.
As shown in the SM~\cite{sm}, qualitatively equivalent results are also obtained for the $B_{2u}$ and $B_{3u}$ pairings.
Here we ignore the insignificant spin-orbit coupling terms, where the appearance of the OCABs in the presence the spin-orbit coupling potentials is demonstrated in the SM~\cite{sm}.
The BdG Hamiltonian $\check{H}_{\boldsymbol{k}}$ belongs to the DIII symmetry class.
\begin{figure}[t]
\begin{center}
\includegraphics[width=0.45\textwidth]{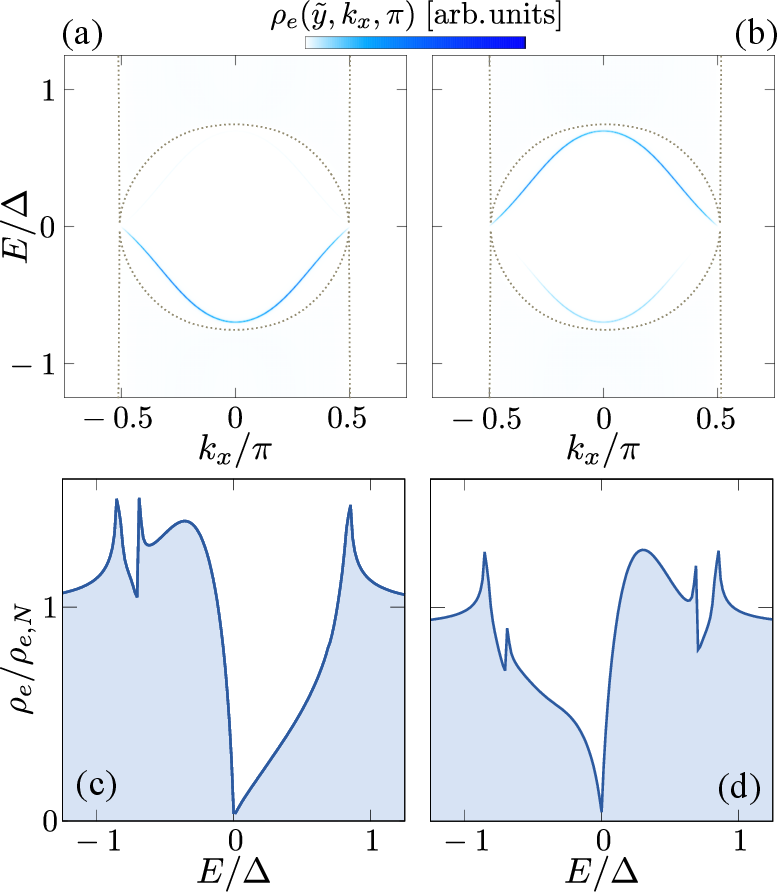}
\caption{\label{fig:2}%
(a)-(b) Electron part of angle-resolved spectral function at $k_z=\pi$ as a function of $k_x$ and the energy $E$.
(c)-(d) Electron part of spectral function as a function of the energy $E$, where the contributions of different $k_x$ and $k_z$ are integrated.
In (a) and (c), we show the results for the outermost site, i.e., $\tilde{y}=1$,
and in (b) and (d), we presents the results for the second outermost site, i.e., $\tilde{y}=1+\frac{1}{2}$.
}
\end{center}
\end{figure}

Although the Hamiltonian $H_{\boldsymbol{k}}$ has the form of a complex $8 \times 8$ matrix,
the energy bands exhibit a two-fold degeneracy at the Brillouin zone boundary, specifically at $k_z=\pi$, due to the sublattice structure.
Thus, we can analytically construct a low-energy effective Hamiltonian for $k_z=\pi$,
which displays the equivalence between the model of UTe$_2$ in Eq.~(\ref{eq:ute2_hm1}) and our minimal model in Eq.~(\ref{eq:minimal_hm}).  
To achieve this, by using a unitary transformation, we deform $H_{\boldsymbol{k}}$ at $k_z=\pi$ into a band basis (see the SM~\cite{sm} for details):
\begin{align}
&\bar{H}^{\mathrm{band}}_{\boldsymbol{q}}=\left[\begin{array}{cc} \check{H}_{\boldsymbol{q},+} & 0 \\ 0 & \check{H}_{\boldsymbol{q},-} \end{array}\right]\nonumber\\
&\check{H}_{\boldsymbol{q},s}=\left[ \begin{array}{cccc}
\epsilon_{\boldsymbol{q}}+sV_{\boldsymbol{q}} & 0 & s\Delta \alpha_{\boldsymbol{q},-} & -s\Delta \beta_{\boldsymbol{q}}\\
0 & \epsilon_{\boldsymbol{q}}-sV_{\boldsymbol{q}} & s\Delta \beta_{\boldsymbol{q}} & s\Delta \alpha_{\boldsymbol{q},+}\\
s\Delta \alpha_{\boldsymbol{q},+} & s\Delta \beta_{\boldsymbol{q}} & -\epsilon_{\boldsymbol{q}}-sV_{\boldsymbol{q}} & 0\\
-s\Delta \beta_{\boldsymbol{q}} & s\Delta \alpha_{\boldsymbol{q},-} & 0 & -\epsilon_{\boldsymbol{q}}+sV_{\boldsymbol{q}}\\
\end{array}\right],\nonumber\\
&V_{\boldsymbol{q}}=\sqrt{m^2_g+f^2_{z,\boldsymbol{q}}},\nonumber\\
&\alpha_{\boldsymbol{q},\pm}=e^{\pm i\frac{k_x}{2}}e^{\pm i\frac{k_y}{2}}\frac{f_{z,\boldsymbol{q}}}{V_{\boldsymbol{q}}},\quad
\beta_{\boldsymbol{q}}=\frac{m_g}{V_{\boldsymbol{q}}},
\label{eq:ute2_hm2}
\end{align}
where $\boldsymbol{q}=(k_x,k_y,k_z=\pi)$.
With the present parameter choice, the normal-state of $\epsilon_{\boldsymbol{q}}+V_{\boldsymbol{q}}$ pinches off from the Fermi level,
i.e., $\epsilon_{\boldsymbol{q}}+V_{\boldsymbol{q}}>0$ irrespective of $\boldsymbol{q}$.
Thus, the low-energy excitation is effectively described by the $2\times2$ Hamiltonian,
\begin{align}
\hat{H}^{\mathrm{eff}}_{\boldsymbol{q},s}=\left[ \begin{array}{cc}
\epsilon_{\boldsymbol{q}}-V_{\boldsymbol{q}} & s\Delta \alpha_{\boldsymbol{q},s}\\
s\Delta \alpha^{\ast}_{\boldsymbol{q},s} & -\epsilon_{\boldsymbol{q}}+V_{\boldsymbol{q}} \end{array} \right].
\label{eq:effective_hm}
\end{align}
Interestingly, although we start from the momentum-independent pair potential, the resultant superconducting gap in the band basis ends up with the momentum-dependent inter-sublattice pair potential:
\begin{align}
\Delta \alpha_{\boldsymbol{q},s} \propto e^{si\frac{k_{x}}{2}}e^{si\frac{k_{y}}{2}} \cos (k_{x}/2) \cos (k_{y}/2),
\end{align}
where the inter-sublattice pair potential is attributed to the inter-rung pairing and the inter-sublattice inter-rung hopping terms. 
Remarkably, we can expect the appearance of the OCABSs because for a fixed $k_{x(y)}$, the effective Hamiltonian in Eq.~(\ref{eq:effective_hm}) is equivalent to Eq.~(\ref{eq:minimal_hm}). 
To verify this expectation, we examine the electron part of the angle-resolved spectral function $\rho_{e}(\tilde{y},k_x,k_z,E)$ at $k_z=\pi$,
where $\tilde{y}=y$ or $y+\frac{1}{2}$ denotes the distance from the surface with $y$ being an integer.
In Fig.~\ref{fig:2}(a) and \ref{fig:2}(b), we show $\rho_{e}(\tilde{y},k_x,\pi,E)$ for the outermost site (i.e., $\tilde{y}=1$) and that for the second outermost site (i.e., $\tilde{y}=1+\frac{1}{2}$)
as a function of $k_x$ and energy $E$.
The Green's function is computed using the recursive Green's function techniques~\cite{fisher_81,ando_91}.
We set the energy broadening as $\delta=10^{-3}\Delta$.
The dotted lines denote the bulk dispersion, obtained by diagonalizing $\bar{H}_{\boldsymbol{q}}$ in Eq.~(\ref{eq:ute2_hm1}).
For $|k_x|<0.5\pi$, we indeed find the ABSs having energies below the superconducting gap.
Moreover, $\rho_{e}(\tilde{y},k_x,\pi,E)$ for the outermost (second outermost) site has the significant enhancement only for $E<0$ ($E>0$),
which means that these ABSs exhibit the oscillating charge nature.
Furthermore, we study the electron part of the spectral function with integrating the contributions of different $k_x$ and $k_z$, i.e., $\rho_{e}(\tilde{y},E)=\sum_{k_x,k_z} \rho_{e}(\tilde{y},k_x,k_z,E)$.
In Figs.~\ref{fig:2}(c) and \ref{fig:2}(d), we show $\rho_{e}(\tilde{y},E)$ at $\tilde{y}=1$ and $\tilde{y}=1+\frac{1}{2}$ as a function of $E$, respectively.
For $\rho_{e}(\tilde{y},E)$ at the outermost (second outermost) site, we find the significant hump only for $E<0$ ($E>0$).
Although the low-energy effective Hamiltonian in Eq.~(\ref{eq:effective_hm}) is only valid at $k_z=\pi$,
the significant asymmetry in $\rho_{e}(\tilde{y},E)$ suggests that the appearance of OCABSs is not restricted to $k_z=\pi$.
As a result, we confirm that the signature of the OCABSs remains distinct even when the contributions from different $k_z$ are integrated.

\begin{figure}[t]
\begin{center}
\includegraphics[width=0.25\textwidth]{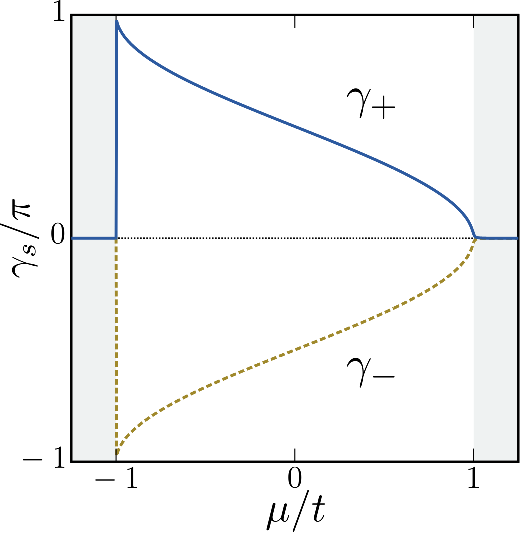}
\caption{\label{fig:3}%
Zak phase of the minima model as a function of the chemical potential.}
\end{center}
\end{figure}

\textit{Discussion}.
Finally, we discuss a significant relationship between the OCABS and a quantum geometry arising from inter-sublattice pairings.
The bulk polarization is related to \textit{unquantized} Zak phases (or Berry phases)~\cite{berry_84,zak_89}, as described by the modern theory of polarization.
At the surface, the  \textit{unquantized} Zak phase leads to charge accumulation, whereby the surface charge is defined only up to an integer multiple of $2\pi$%
~\cite{king-smith_93,vanderbilt_93,resta_94,resta_07,bardarson_17,ortix_17,watanabe_18}
(see also Refs.~\cite{rhim_20,pletyukhov_20} for alternative characterizations of surface topology).
The Rice--Mele model hosts in-gap surface states that contribute significantly to the probability density of the surface charge.
In superconductors, the Zak phase is intrinsically quantized to zero or $\pi$ due to particle-hole symmetry.
Consequently, the concept of \textit{unquantized} Zak phases has, to the best of our knowledge, not been discussed in the context of superconductors.
Nevertheless, the constraint from the particle-hole symmetry can be relaxed within block components: while particle-hole symmetry is satisfied for the full Hamiltonian, each block component breaks it, thereby allowing the definition of a unquantized Zak phase for each block component. The BdG Hamiltonian of our minimal model is a representative example hosting the unquantized Zak phase.
Here, we compute the Zak phase defined for each of $\hat{H}_{k,s}$:
\begin{align}
\gamma_s=i \int^{\pi}_{-\pi} dk \braket{u_{k,s}| \partial_k  |u_{k,s}},
\end{align}
where $\ket{u_{k,s}}$ describes the occupied eigenstate of $\hat{H}_{k,s}$ and satisfies the periodic boundary condition $\ket{u_{k,s}} =\ket{u_{k+2\pi,s}}$
\footnote{In this paper, we focus only on the intercellular part of the Zak phase~\cite{bardarson_17}.}.
In Fig.~\ref{fig:3}, we show $\gamma_s$ as a function of $\mu$, where $\Delta=0.001t$.
For $|\mu|<t$, which is the parameter region hosting the OCABSs, we find that the Zak phase $\gamma_s$ takes unquantized values between $-\pi$ and $\pi$.
The particle-hole symmetry of the full BdG Hamiltonian $\check{H}$ imposes $\gamma_+ +\gamma_- = 0 \mod 2 \pi$, indicating that our minimal model is topologically trivial.
Further investigation into the relationship between unquantized Zak phases and OCABSs in a system beyond the Rice--Mele model can be a fruitful future research.

Investigating the appearance of OCABSs in other sublattice superconductors is also an interesting future work.
Based on our results, we suggest that not only the sublattice degrees of freedom but also the inter-sublattice pairings are essential for the appearance of OCABSs.
There are a number of superconducting systems with the sublattice degree of freedom, such as iron-based superconductors~\cite{hu_13,vafek_13,hu_14,luo_23,xue_24}
and superconducting ladder/layer systems%
~\cite{scalapino_92,sigrist_93,rice_96,kinoshita_96,arita_05,akimitsu_06,ohgushi_15,kuroki_20,scalettar_92,scalettar_94,hanke_94,arita_02,scalapino_11,kuroki_17,wang_23,kuroki_23},
while it remains to be clarified whether inter-sublattice pairings are realized in these systems.
Moreover, proposing conclusive experiments to observe the oscillating charge nature of the OCABSs is a desirable future task.
A tunneling spectroscopy experiment on the superconductor UTe$_2$ recently observed distinctly asymmetric conductance spectra~\cite{madhavan_20}.
It would be important to verify whether this phenomenon is caused by the OCABSs.

In summary, we have demonstrated the appearance of unconventional ABSs with exotic charge characteristics, i.e., the OCABSs.
The complete breakdown of the proportionality in $\rho_e$, $\rho$, and $G$ is a remarkable consequence of the charge oscillating nature.
We have also described the possible emergence of OCABSs in the explicit model of UTe$_2$ and locally noncentrosymmetric superconductors.
Our results indicate a new direction for the development of the physics of ABSs.

\begin{acknowledgments}
We thank Y. Tanaka and L. Katayama for the insightful discussions. 
S.A. is supported by the Nagoya University Interdisciplinary Frontier Fellowship (Grant No.JPMJFS2120).
S.I. is supported by the Grant-in-Aid for JSPS Fellows (JSPS KAKENHI Grant No. JP22KJ1507) and for Early-Career Scientists (JSPS KAKENHI Grant No. JP24K17010).
\end{acknowledgments}

\pagebreak
\onecolumngrid
\begin{center}
  \textbf{\large Supplemental Material for ``Oscillating-charged Andreev Bound States and Their Appearance in UTe$_2$''}\\ \vspace{0.3cm}
Satoshi Ando$^{1}$, Shingo Kobayashi$^{2}$, Andreas P. Schnyder$^{3}$, Yasuhiro Asano$^{4}$, and Satoshi Ikegaya$^{1,5}$\\ \vspace{0.1cm}
{\itshape $^{1}$Department of Applied Physics, Nagoya University, Nagoya 464-8603, Japan\\
$^{2}$RIKEN Center for Emergent Matter Science, Wako, Saitama, 351-0198, Japan\\
$^{3}$Max-Planck-Institut f\"ur Festk\"orperforschung, Heisenbergstrasse 1, D-70569 Stuttgart, Germany\\
$^{4}$Department of Applied Physics, Hokkaido University, Sapporo 060-8628, Japan\\
$^{5}$Institute for Advanced Research, Nagoya University, Nagoya 464-8601, Japan}
\date{\today}
\end{center}

\section{Oscillating-charged Andreev bound states \\ in locally noncentrosymmetric superconductors}
\begin{figure}[t]
\begin{center}
\includegraphics[width=0.4\textwidth]{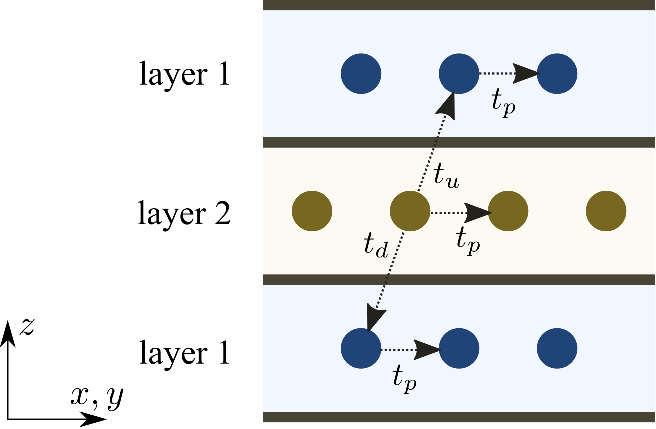}
\caption{\label{fig:sm1}%
Schematic image of the hopping terms in the locally noncentrosymmetric superconductor.}
\end{center}
\end{figure}

In this section, we show the appearance of the oscillating-charged Andreev bound state (OCABS) in a locally noncentrosymmetric superconductor with an interlayer pairing.
The Bogoliubov--de Gennes (BdG) Hamiltonian of the locally noncentrosymmetric superconductor, which is proposed in Refs.~\cite{ramires_21(1),ramires_21(2)}, is given by
\begin{align}
\begin{split}
&H = \frac{1}{2}\sum_{\boldsymbol{k},\sigma}\Psi^{\dagger}_{\boldsymbol{k}\sigma} \bar{H_{\boldsymbol{k}}} \Psi^{\dagger}_{\boldsymbol{k}\sigma}+ H_{\mathrm{soc}},\\
&\Psi^{\dagger}_{\boldsymbol{k}\sigma}=[\psi^{\dagger}_{\boldsymbol{k}\sigma},\psi^{\mathrm{T}}_{-\boldsymbol{k}\bar{\sigma}}],\quad
\psi^{\dagger}_{\boldsymbol{k}\sigma}=[\psi^{\dagger}_{\boldsymbol{k},\sigma,a,1},\psi^{\dagger}_{\boldsymbol{k},\sigma,a,2},
\psi^{\dagger}_{\boldsymbol{k},\sigma,b,1},\psi^{\dagger}_{\boldsymbol{k},\sigma,b,2}],\\
&\bar{H_{\boldsymbol{k}}}=\left[\begin{array}{cc}\check{h}({\boldsymbol{k}}) & \check{h}_\Delta({\boldsymbol{k}}) \\
\check{h}^{\dagger}_\Delta({\boldsymbol{k}}) & -\check{h}^{\ast}(-{\boldsymbol{k}}) \end{array}\right],\quad
\check{h}({\boldsymbol{k}})=
\left[\begin{array}{cc} \hat{K}_{\boldsymbol{k}} & \hat{T}^{\dagger}_{\boldsymbol{k}} \\
\hat{T}_{\boldsymbol{k}} & \hat{K}_{\boldsymbol{k}} \end{array} \right],\quad
\check{h}_{\Delta}({\boldsymbol{k}})=\left[\begin{array}{cc} 0 & \hat{\Delta}({\boldsymbol{k}}) \\ \hat{\Delta}^{\dagger}({\boldsymbol{k}}) & 0 \end{array} \right], \\
&\hat{K}_{\boldsymbol{k}}=\left[ \begin{array}{cc} h_{00}(\boldsymbol{k}) & 0 \\ 0 & h_{00}(\boldsymbol{k}) \end{array}\right],\quad
\hat{T}_{\boldsymbol{k}}=e^{i\frac{k_x}{2}}e^{i\frac{k_y}{2}}e^{i\frac{k_z}{2}}
\left[ \begin{array}{cc} 0 & h_{10}(\boldsymbol{k})-ih_{20}(\boldsymbol{k}) \\ h_{10}(\boldsymbol{k})+ih_{20}(\boldsymbol{k}) & 0 \end{array}\right],\\
&\hat{\Delta} = e^{i\frac{k_x}{2}}e^{i\frac{k_y}{2}}e^{i\frac{k_z}{2}} \left[ \begin{array}{cc} 0 & d_{13}({\boldsymbol{k}}) \\ d_{13}({\boldsymbol{k}}) & 0 \end{array} \right],\\
&h_{00}(\boldsymbol{k})=2t_p \left[ \cos (k_x) + t_2 \cos (k_y) \right] -\mu,\\
&h_{10}(\boldsymbol{k})=4(t_u+t_d) \cos(k_x/2)\cos(k_y/2)\cos(k_z/2),\\
&h_{20}(\boldsymbol{k})=-4(t_u-t_d) \cos(k_x/2)\cos(k_y/2)\sin(k_z/2),\\
&d_{13}(\boldsymbol{k})= \Delta \cos(k_x/2)\cos(k_y/2)\sin(k_z/2),\\
&H_{\mathrm{soc}}=\frac{1}{2}\sum_{\boldsymbol{k}}
[\psi^{\dagger}_{\boldsymbol{k}\uparrow},\psi^{\dagger}_{\boldsymbol{k}\downarrow},
\psi^{\mathrm{T}}_{-\boldsymbol{k}\uparrow},\psi^{\mathrm{T}}_{-\boldsymbol{k}\downarrow}]
\tilde{\Lambda}_{\boldsymbol{k}}
\left[\begin{array}{cc}\psi_{\boldsymbol{k}\uparrow}\\ \psi_{\boldsymbol{k}\downarrow} \\
\psi^{\ast}_{-\boldsymbol{k}\uparrow}\\ \psi^{\ast}_{-\boldsymbol{k}\downarrow}\end{array}\right],\\
&\tilde{\Lambda}_{\boldsymbol{k}}=\left[\begin{array}{cc}\bar{\lambda}_{\boldsymbol{k}} & 0 \\
0 & -\bar{\lambda}^{\ast}_{-\boldsymbol{k}} \end{array}\right],\quad
\bar{\lambda}_{\boldsymbol{k}} =
\left[\begin{array}{cc}\check{\lambda}_{33}(\boldsymbol{k}) & \check{\lambda}_{31}(\boldsymbol{k})-i\check{\lambda}_{32}(\boldsymbol{k}) \\
\check{\lambda}_{31}(\boldsymbol{k})+i\check{\lambda}_{32}(\boldsymbol{k}) & -\check{\lambda}_{33}(\boldsymbol{k}) \end{array}\right],\\
&\check{\lambda}_{3\nu}(\boldsymbol{k})=
\left[\begin{array}{cc}\hat{\lambda}_{3\nu}(\boldsymbol{k}) & 0 \\ 0 & \hat{\lambda}_{3\nu}(\boldsymbol{k}) \end{array}\right], \quad
\hat{\lambda}_{3\nu}(\boldsymbol{k})=
\left[\begin{array}{cc}h_{3\nu}(\boldsymbol{k}) & 0 \\ 0 & -h_{3\nu}(\boldsymbol{k}) \end{array}\right],\\
&h_{31}(\boldsymbol{k}) = -\alpha \sin(k_y), \quad
h_{32}(\boldsymbol{k}) = \alpha \sin(k_x),\quad
h_{33}(\boldsymbol{k})=\lambda \sin(k_x) \sin(k_y) \sin(k_z) [\cos (k_x) - \cos(k_y)],
\label{eq:cerh2as2_hm0}
\end{split}
\end{align}
where $\psi_{\boldsymbol{k},\sigma,\alpha,l}$ is the annihilation operator of an electron with momentum $\boldsymbol{k}$ and spin $\sigma$ at the $l$-th layer of the sublattice $\alpha$.
The index $\bar{\sigma}$ denotes the opposite spin of $\sigma$.
Each hoping term is shown schematically in Fig.~\ref{fig:sm1}.
The difference between $t_u$ and $t_d$ reflects the absence of local inversion symmetry.
Specifically, the term of $h_{20}$ in the Hamiltonian, which is an odd function with respect to $k_z$, becomes finite for $t_u \neq t_d$.
The strength of the Rashba (Ising-type) spin-orbit coupling are described by $\alpha$ ($\lambda$).
Note that Eq.~(\ref{eq:cerh2as2_hm0}) is modified to explicitly describe the sublattice degrees of freedom.
We assume the inter-layer-even spin-triplet odd-parity pair potential~\cite{ramires_21(1),ramires_21(2)}.
When we neglect the interlayer hopping terms and the spin-orbit coupling terms, i.e.,
\begin{align}
h_{10}(\boldsymbol{k})=h_{20}(\boldsymbol{k})=h_{31}(\boldsymbol{k})=h_{32}(\boldsymbol{k})=h_{33}(\boldsymbol{k}) = 0,
\end{align}
the BdG Hamiltonian is rewritten as
\begin{align}
\begin{split}
& H=\frac{1}{2}\sum_{\boldsymbol{k},\sigma,l} [\Psi^{\dagger}_{\boldsymbol{k},\sigma,l,+},\Psi^{\dagger}_{\boldsymbol{k},\sigma,l,-}]
\left[\begin{array}{cc}\hat{h}_{\boldsymbol{k},+}& 0 \\ 0 & \hat{h}_{\boldsymbol{k},-}\end{array}\right]
\left[\begin{array}{cc}\Psi_{\boldsymbol{k},\sigma,l,+}\\\Psi_{\boldsymbol{k},\sigma,l,-}\end{array}\right],\\
&\hat{h}_{\boldsymbol{k},s}=\left[\begin{array}{cc}t_p \cos (k_x) - \mu(k_y) & \Delta_s(k_y,k_z) e^{-is \frac{k_x}{2}} \cos (k_x/2) \\
\Delta_s^{\ast}(k_y,k_z) e^{is \frac{k_x}{2}} \cos (k_x/2) & -t_p \cos (k_x) + \mu(k_y)\end{array}\right], \\
&\mu(k_y) = \mu - t_p \cos(k_y), \quad
\Delta_s(k_y,k_z) = \Delta e^{-is \frac{k_y}{2}} e^{-is \frac{k_z}{2}} \cos (k_y/2) \sin(k_z/2),\\
&\Psi_{\boldsymbol{k},\sigma,l,+}=[\psi_{\boldsymbol{k},\sigma,a,l},\psi^{\dagger}_{-\boldsymbol{k},\sigma,b,\bar{l}}]^{\mathrm{T}},\quad
\Psi_{\boldsymbol{k},\sigma,l,-}=[\psi_{\boldsymbol{k},\sigma,b,l},\psi^{\dagger}_{-\boldsymbol{k},\sigma,a,\bar{l}}]^{\mathrm{T}},
\label{eq:minimal_hm}
\end{split}
\end{align}
where $s=+1$ ($-1$) and $\bar{l}$ denotes the opposite layer of $l$.
With fixed $k_y$ and $k_z$, $\hat{h}_{\boldsymbol{k},s}$ coincides with the BdG Hamiltonian of the minimal model in Eq.~(1) of the main text.
Therefore, we can expect the appearance of the OCABSs in the locally noncentrosymmetric superconductor with the interlayer pairing.

\begin{figure}[h]
\begin{center}
\includegraphics[width=0.5\textwidth]{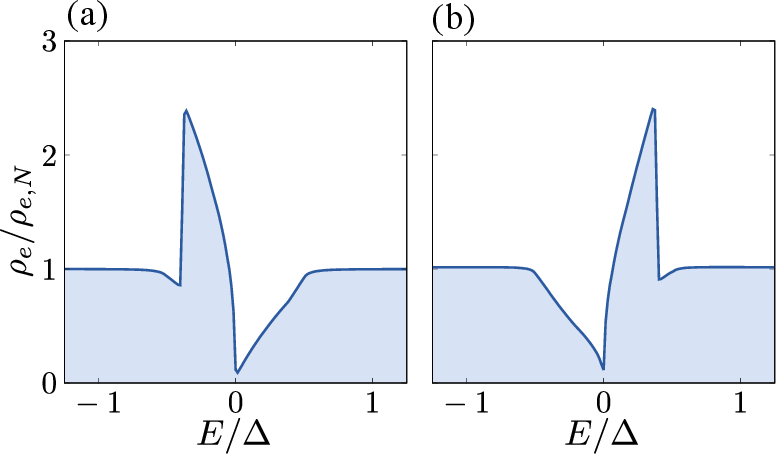}
\caption{\label{fig:sm2}%
Electron part of the spectral function in the case of the dominant interlayer hopping (i.e., $t_u,t_d>\alpha$) as a function of the energy $E$.
In (a) and (b), we show the result for the outermost site, i.e., $\tilde{x}=1$, and that for the second outermost site, i.e., $\tilde{x}=1+\frac{1}{2}$, respectively.
}
\end{center}
\end{figure}

\begin{figure}[h]
\begin{center}
\includegraphics[width=0.5\textwidth]{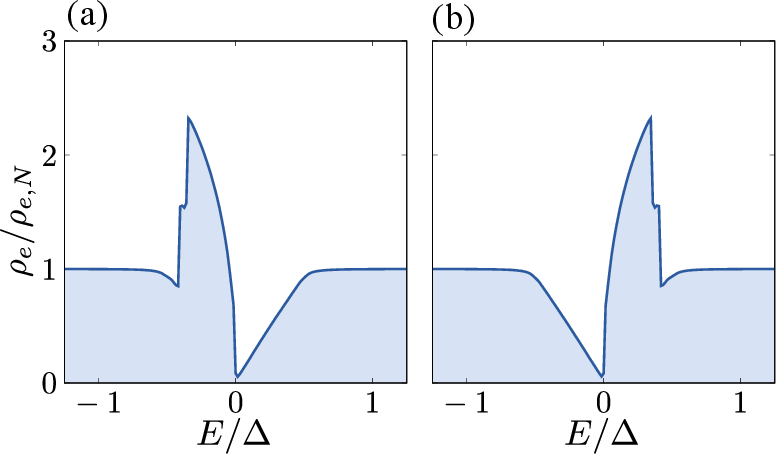}
\caption{\label{fig:sm3}%
Electron part of the spectral function in the case of the dominant Rashba spin-orbit coupling (i.e., $t_u,t_d<\alpha$) as a function of the energy $E$.
In (a) and (b), we show the result for the outermost site, i.e., $\tilde{x}=1$, and that for the second outermost site, i.e., $\tilde{x}=1+\frac{1}{2}$, respectively.
}
\end{center}
\end{figure}

To demonstrate the appearance of the OCABSs, we examine the electron part of the spectral function, which is defined by
\begin{align}
&\rho_{e}(\tilde{x},E)=\sum_{k_y,k_z} \rho_{e}(\tilde{x},k_y,k_z,E),\quad
\rho_e(\tilde{x},k_y,k_z,E)=-\frac{1}{\pi}\mathrm{Im}[g_{k_y,k_z}(\tilde{x},\tilde{x},E+i\delta)],
\end{align}
where $\tilde{x}=x$ or $x+\frac{1}{2}$ measures the distance from the surface with $x$ being an integer,
and $g_{k_y,k_z}(\tilde{x},\tilde{x},E)$ is the retarded Green's function obtained using the recursive Green's function techniques~\cite{fisher_81,ando_91}.
The parameter $\delta$, which characterizes the broadening of the energy, is fixed at $\delta=10^{-3}\Delta$.
In Figs.~\ref{fig:sm2}(a) and \ref{fig:sm2}(b), we show $\rho_{e}(\tilde{x},E)$ 
at the outermost site ($\tilde{x}=1$) and the second outermost site ($\tilde{x}=1+\frac{1}{2})$ as a function of the energy, respectively.
We consider the dominant interlayer hopping case, i.e., $t_u,t_d>\alpha$ with $(\mu,$ $t_p,$ $t_u,$ $t_d,$ $\alpha,$ $\lambda,$ $\Delta)$$=$$(0,$ $1,$ $0.2,$ $0.2,$ $0.05,$ $0,$ $0.001)$.
Here we ignore the Ising-type spin-orbit coupling associated with the interlayer long-ranged hopping.
In Figs.~\ref{fig:sm3}(a) and \ref{fig:sm3}(b), we show $\rho_{e}(\tilde{x},E)$ with the dominant Rashba spin-orbit coupling case, i.e., $t_u,t_d<\alpha$,
with $(\mu,$ $t_p,$ $t_u,$ $t_d,$ $\alpha,$ $\lambda,$ $\Delta)$$=$$(0,$ $1,$ $0.05,$ $-0.05,$ $0.2,$ $0,$ $0.001)$.
In both cases shown in Fig.~\ref{fig:sm2} and Fig.~\ref{fig:sm3}, $\rho_{e}(\tilde{x},E)$ at the outermost (second outermost) has the significant peak only for $E<0$ ($E>0$).
Thus, we confirm the appearance of the OCABSs in the locally noncentrosymmetric superconductor with the interlayer pairing.

\clearpage

\section{Energy eigenvalue and wave function \\ of the oscillating-charged Andreev bound state}
In this section, we calculate the energy eigenvalue and wave function of the OCABS in the minimal model.
We consider the BdG equation, which corresponds to Eq.~(3) in the main text:
\begin{align}
\hat{T}_{s}\varphi_{x-1,s}+\hat{T}^{\dagger}_{s}\varphi_{x+1,s}+\hat{K}\varphi_{x,s} = E_{s}\varphi_{x,s},
\label{eq:bdg1}
\end{align}
with
\begin{align}
\begin{split}
&\hat{T}_{s}=\left[\begin{array}{cc}\frac{t}{2} & \frac{(1+s)\Delta}{4} \\ \frac{(1-s)\Delta}{4} & -\frac{t}{2}\end{array}\right],\quad
\hat{K}=\left[\begin{array}{cc}-\mu & \frac{\Delta}{2} \\ \frac{\Delta}{2} & \mu\end{array}\right],\\
&\varphi_{x,+}=[u_{x,+},v_{x+\frac{1}{2},+}]^{\mathrm{T}}, \; \varphi_{x,-}=[u_{x+\frac{1}{2},-},v_{x,-}]^{\mathrm{T}},\\
&u_{x+\frac{1}{2},+}=v_{x,+}=u_{x,-}=v_{x+\frac{1}{2},-}=0,
\end{split}
\end{align}
where $u_{x,s}$ ($v_{x,s}$) denotes the electron (hole) component in the $a$-sublattice,
and $u_{x+\frac{1}{2},s}$ ($v_{x+\frac{1}{2},s}$) represents the electron (hole) component in the $b$-sublattice.
We consider the semi-infinite system for $x>0$ and apply an open-boundary condition, $\varphi_{0,s}=0$.
We first focus on the solution belonging to $s=+$.
Substituting
\begin{align}
\varphi_{x,+}=\left[\begin{array}{cc} u_{x,+} \\ v_{x+\frac{1}{2},+} \end{array}\right]=\left[\begin{array}{cc} u_{k} \\ v_{k} \end{array}\right]e^{ikx},
\end{align}
into Eq.~(\ref{eq:bdg1}), we obtain:
\begin{align}
\left[\begin{array}{cc} t \cos k -\mu & e^{-ik/2} \Delta \cos(k/2) \\ e^{ik/2} \Delta \cos(k/2) & -t \cos k +\mu \end{array} \right] \left[\begin{array}{cc} u_{k} \\ v_{k} \end{array}\right]
=E \left[\begin{array}{cc} u_{k} \\ v_{k} \end{array}\right].
\label{eq:bdg2}
\end{align}
From Eq.~(\ref{eq:bdg2}), we find
\begin{align}
E = \pm \sqrt{(t \cos (k) -\mu)^2 + \Delta^2 \cos^2(k/2)}, \label{eq:eng1}
\end{align}
and
\begin{align}
\left[\begin{array}{cc} u_{k} \\ v_{k} \end{array}\right] \propto
\left[\begin{array}{cc} E+(t \cos (k) -\mu) \\ e^{ik/2}\Delta \cos(k/2) \end{array}\right] \propto
\left[\begin{array}{cc} e^{-ik/2}\Delta \cos(k/2) \\ E-(t \cos (k) -\mu) \end{array}\right]. \label{eq:wav1}
\end{align}
To investigate Andreev bound states (ABSs), we assume
\begin{align}
k = q + i \kappa \label{eq:momentum}
\end{align}
where $q$ and $\kappa$ are real numbers.
By substituting Eq.~(\ref{eq:momentum}) into Eq.~(\ref{eq:eng1}), we obtain
\begin{align}
E^2 & = t^2(\cos^2 (q) \cosh^2 (\kappa) - \sin^2 (q) \sinh^2 (\kappa)) - 2 t \mu \cos (q) \cosh (\kappa) + \mu^2 + \Delta^2 \frac{1+\cos (q) \cosh (\kappa)}{2} \nonumber\\
& + i \left(-2t^2 \cos (q) \cosh (\kappa) \sin (q) \sinh (\kappa) + 2t \mu \sin (q) \sinh (\kappa) - \Delta^2 \frac{\sin (q) \sinh (\kappa)}{2} \right).
\label{eq:eng2}
\end{align}
From the imaginary part of Eq.~(\ref{eq:eng2}), we obtain
\begin{align}
\cos (q) \cosh (\kappa) = \frac{4 t \mu - \Delta^2}{4 t^2}.  \label{eq:eng3_i}
\end{align}
By using the real part of Eq.~(\ref{eq:eng2}) and Eq.~(\ref{eq:eng3_i}), we obtain
\begin{align}
\sin^2 (q) \sinh^2 (\kappa) = \frac{8 t \mu \Delta^2 - \Delta^4 + 8 t^2 \Delta^2 - 16 t^2 E^2}{16 t^4}. \label{eq:eng3_r}
\end{align}
Here we assume $\Delta/t \ll 1$ and approximate Eq.~(\ref{eq:eng3_i}) and Eq.~(\ref{eq:eng3_r}) as
\begin{align}
&\cos (q) \cosh (\kappa) = \frac{\mu}{t},  \label{eq:eng4_i}\\
&\sin^2 (q) \sinh^2 (\kappa) = \frac{t \mu \Delta^2 + t^2 \Delta^2 - 2 t^2 E^2}{2 t^4}, \label{eq:eng4_r}
\end{align}
respectively.
In general, the decay length of the ABS increases (i.e., $\kappa$ decreases) by decreasing $\Delta$.
Thus, in the limit of $\Delta/t \ll 1$, we can expect $\kappa \ll 1$.
Based on this expectation, we approximate Eq.~(\ref{eq:eng4_i}) and Eq.~(\ref{eq:eng4_r}) as
\begin{align}
&\cos (q)  = \frac{\mu}{t},  \label{eq:eng5_i}\\
&\kappa^2 \sin^2 (q) = \frac{t \mu \Delta^2 + t^2 \Delta^2 - 2 t^2 E^2}{2 t^4}, \label{eq:eng5_r}
\end{align}
respectively.
The expectation of $\kappa \ll 1$ is justified later.
From Eq.~(\ref{eq:eng5_i}), we obtain
\begin{align}
q = \pm q_F, \quad q_F = \arccos \left( \frac{\mu}{t} \right), \label{eq:eng6_i}
\end{align}
where $-1 < \mu/t < 1$ is satisfied.
From Eq.~(\ref{eq:eng5_r}) and Eq.~(\ref{eq:eng6_i}), we obtain
\begin{align}
\kappa = \pm \kappa_E, \quad \kappa_E = \sqrt{\frac{t \mu \Delta^2 + t^2 \Delta^2 - 2 t^2 E^2}{2 t^2 (t^2-\mu^2)}}.
\label{eq:eng6_r}
\end{align}
As a result, the wave function of the ABS is represented by
\begin{align}
\begin{split}
&\varphi_{x,+} = a \left[\begin{array}{cc} E+(t \cos (k_+) -\mu) \\ e^{ik_+/2}\Delta \cos(k_+/2) \end{array}\right] e^{i k_+ x}
+ b \left[\begin{array}{cc} e^{ik_-/2}\Delta \cos(k_-/2) \\ E-(t \cos (k_-) -\mu) \end{array}\right] e^{-i k_- x}, \label{eq:wav2}\\
& k_{\pm}=q_F \pm i\kappa_E
\end{split}
\end{align}
where $a$ and $b$ are numerical coefficients.
The wave function of $\varphi_{x,+}$ satisfies
\begin{align}
\lim_{x\rightarrow\infty}\varphi_{x,+} = 0.
\end{align}
From the conditions of $\Delta/t \ll 1$ and $\kappa \ll 1$, we can approximate
\begin{align}
&\varphi_{x,+} \approx a \left[\begin{array}{cc} E-it \sin (q_F) \kappa_E \\ e^{iq_F}\Delta \cos(q_F/2) \end{array}\right] e^{i k_+ x}
+ b \left[\begin{array}{cc} e^{iq_F/2}\Delta \cos(q_F/2) \\ E-it \sin (q_F) \kappa_E \end{array}\right] e^{-i k_- x}. \label{eq:wav3}
\end{align}
From the boundary condition of $\varphi_{x=0,+}=0$, we obtain
\begin{align}
\left[\begin{array}{cc} E-it \sin (q_F) \kappa_E & e^{iq_F/2}\Delta \cos(q_F/2) \\ e^{iq_F/2}\Delta \cos(q_F/2) & E-it \sin (q_F) \kappa_E  \end{array}\right]
\left[\begin{array}{cc} a \\ b \end{array}\right] = 0, \label{eq:wav4}
\end{align}
which leads
\begin{align}
\left|\begin{array}{cc} E-it \sin (q_F) \kappa_E & e^{iq_F/2}\Delta \cos(q_F/2) \\ e^{iq_F/2}\Delta \cos(q_F/2) & E-it \sin (q_F) \kappa_E  \end{array}\right|
= 0. \label{eq:wav5}
\end{align}
By solving Eq.~(\ref{eq:wav5}), we obtain the energy eigenvalue of the ABS as
\begin{align}
E = E_+ = - \Delta \frac{t+\mu}{2t}. \label{eq:eng7}
\end{align}
By substituting Eq.~(\ref{eq:eng7}) into Eq.~(\ref{eq:eng6_r}), we find
\begin{align}
\kappa_E = \kappa_0 = \frac{\Delta}{2t},
\end{align}
which satisfies $\kappa_0 \ll 1$ for $\Delta/t \ll 1$.
By substituting Eq.~(\ref{eq:eng7}) into Eq.~(\ref{eq:wav4}), we obtain
\begin{align}
\left[\begin{array}{cc} 1& -1 \\ -1 & 1  \end{array}\right]
\left[\begin{array}{cc} a \\ b \end{array}\right] = 0, \label{eq:wav6}
\end{align}
which leads $a=b$.
As a result, we find
\begin{align}
&\varphi_{x,+} = a \left( \left[\begin{array}{cc} 1 \\ -1 \end{array}\right] e^{i k_+ x}
+ \left[\begin{array}{cc} -1 \\ 1 \end{array}\right] e^{-i k_- x}\right) 
=2ia \left[\begin{array}{cc} 1 \\ -1 \end{array}\right] \sin (q_F x) e^{-\kappa_0 x}. \label{eq:wav7}
\end{align}
From the normalization condition,
\begin{align}
\sum_x |\varphi_{x,+}|^2 = 1,
\end{align}
we obtain
\begin{align}
\varphi_{x,+}=\left[\begin{array}{cc} 1 \\ -1 \end{array}\right] \phi_x, \quad
\phi_x = \sqrt{2 \kappa_0} \sin (q_F x) e^{-\kappa_0 x}. \label{eq:wav8}
\end{align}
The energy eigenvalue in Eq.~(\ref{eq:eng7}) and wave function in Eq.~(\ref{eq:wav8}) correspond to those in Eq.~(5) in the main text.
Similarly, the energy eigenvalue and wave function of the OCABS for $s=-$ are obtained as
\begin{align}
E_- = \Delta \frac{t+\mu}{2t}, \quad \varphi_{x,-}=\left[\begin{array}{cc} 1 \\ 1 \end{array}\right] \phi_x,
\label{eq:ocabs2}
\end{align}
respectively.

\section{The detailed derivation for the physical quantities}
In this section, we provide a detailed derivation for the following three quantities in the presence of an ABS:
the electron part of the spectral function $\rho_e$, the local density of states $\rho$, and the tunneling conductance spectrum $G$.
In the following, we assume that the ABS exists at an energy $E_B$, which is energetically well separated from other states,
and the wave function of the ABS is represented by $\varphi^B_{\tilde{x}}=[u^B_{\tilde{x}},v^T _{\tilde{x}}]^{\mathrm{T}}$.

\subsection{Electron part of the spectral function and Local density of states}
We first examine $\rho_e$ and $\rho$, which are given by
\begin{align}
\begin{split}
&\rho_e(\tilde{x},E)=-\frac{1}{\pi}\mathrm{Im}[g(\tilde{x},\tilde{x},E+i\delta)], \\
&\rho(\tilde{x},E)=-\frac{1}{\pi}\mathrm{Im}[g(\tilde{x},\tilde{x},E+i\delta)+\underline{g}(\tilde{x},\tilde{x},E+i\delta)], \label{eq:rho_def}
\end{split}
\end{align}
where
\begin{align}
\mathcal{G}(\tilde{x},\tilde{x}^{\prime},E)=\left[ \begin{array}{cc}
g(\tilde{x},\tilde{x}^{\prime},E)&f(\tilde{x},\tilde{x}^{\prime},E) \\
\underline{f}(\tilde{x},\tilde{x}^{\prime},E)&\underline{g}(\tilde{x},\tilde{x}^{\prime},E)\end{array} \right],
\end{align}
represents the Gorkov--Green's function, and $\delta$ characterizes the broadening of the energy.
In general, the electron Green's function and hole Green's function are given by
\begin{align}
\begin{split}
&g(\tilde{x},\tilde{x},E) = \sum_{\nu}\left[ \frac{|u_{\nu}(\tilde{x})|^2}{E-E_{\nu}+i\delta} + \frac{|v_{\nu}(\tilde{x})|^2}{E+E_{\nu}+i\delta} \right],\\
&\underline{g}(\tilde{x},\tilde{x},E) = \sum_{\nu}\left[ \frac{|v_{\nu}(\tilde{x})|^2}{E-E_{\nu}+i\delta} + \frac{|u_{\nu}(\tilde{x})|^2}{E+E_{\nu}+i\delta} \right],
\end{split}
\end{align}
respectively, where $u_{\nu}(x)$ ($v_{\nu}(x)$) represents the electron (hole) components of the wave function belonging to the energy eigenvalue $E_{\nu}$.
Since we assume $|E_B - E_{\nu}|\gg1$ for $\nu \neq B$, the Green's functions at $E \approx E_B$ are obtained by
\begin{align}
\begin{split}
&g(\tilde{x},\tilde{x},E) \approx \frac{n^e_{\tilde{x}}}{E-E_B+i\delta},\\
&\underline{g}(\tilde{x},\tilde{x},E) \approx \frac{n^h_{\tilde{x}}}{E-E_B+i\delta}, \label{eq:green_abs}
\end{split}
\end{align}
where  $n^e_{\tilde{x}}=|u^B_{\tilde{x}}|^2$ and $n^h_{\tilde{x}}=|v^B_{\tilde{x}}|^2$.
By substituting Eq.~(\ref{eq:green_abs}) into Eq.~(\ref{eq:rho_def}), we obtain
\begin{align}
\begin{split}
& \rho_e(\tilde{x},E) = \frac{1}{\pi}\frac{\delta}{(E-E_B)^2 + \delta^2} n^e_{\tilde{x}},\\
& \rho(\tilde{x},E) = \frac{1}{\pi}\frac{\delta}{(E-E_B)^2 + \delta^2} n_{\tilde{x}}, \label{rho_abs}
\end{split}
\end{align}
where $n_{\tilde{x}}=n^e_{\tilde{x}}+n^h_{\tilde{x}}$.
The equation~(\ref{rho_abs}) corresponds to Eq.~(11) in the main text.

\subsection{Tunneling conductance}
In this section, we derive the tunneling conductance given by Eq.~(13) in the main text.
We start with a tight-binding BdG Hamiltonian describing a normal-metal--superconductor junction,
\begin{align}
\begin{split}
&H = H_S + H_N + H_T,\\
&H_S = \sum_{\tilde{x},\tilde{x}^{\prime}>0}
[c^{\dagger}_{\tilde{x}}, c_{\tilde{x}}]
\left[ \begin{array}{cc} h(\tilde{x},\tilde{x}^{\prime}) & \Delta(\tilde{x},\tilde{x}^{\prime}) \\
-\Delta^{\ast}(\tilde{x},\tilde{x}^{\prime}) & -h^{\ast}(\tilde{x},\tilde{x}^{\prime}) \end{array} \right]
\left[ \begin{array}{cc} c_{\tilde{x}^{\prime}}\\ c^{\dagger}_{\tilde{x}^{\prime}}\end{array} \right],\\
&H_N = -t\sum_{x<0} (a^{\dagger}_{x+1} a_x + \mathrm{h.c.}),\\
&H_T = -t_T (c^{\dagger}_{\tilde{x}=1} a_{x=0} + \mathrm{h.c.}),
\end{split}
\end{align}
where $c_{\tilde{x}}$ ($a_x$) represents the annihilation operator of the electron in the superconducting (normal) segment.
For the superconducting segment (i.e., $\tilde{x} >0$), the kinetic energy in the normal state is described by $h(\tilde{x},\tilde{x}^{\prime})$,
and the pair potential is given by $\Delta(\tilde{x},\tilde{x}^{\prime})$.
The hopping integral in the normal segment ($x \leq0$) is given by $t$.
The hopping integral at the interface is represented by $t_T$.
The Bogoliubov transformation is denoted by
\begin{align}
\left[\begin{array}{cc} c_{\tilde{x}}\\ c^{\dagger}_{\tilde{x}}\end{array} \right]
=\sum_{\nu} \left( \left[\begin{array}{cc} u_{\nu}(\tilde{x})\\ v_{\nu}(\tilde{x})\end{array} \right] \gamma_{\nu}
+ \left[\begin{array}{cc} v^{\ast}_{\nu}(\tilde{x})\\ u^{\ast}_{\nu}(\tilde{x})\end{array} \right] \gamma^{\dagger}_{\nu}\right),
\end{align}
where
\begin{align}
\sum_{\tilde{x}^{\prime}>0}
\left[ \begin{array}{cc} h(\tilde{x},\tilde{x}^{\prime}) & \Delta(\tilde{x},\tilde{x}^{\prime}) \\
-\Delta^{\ast}(\tilde{x},\tilde{x}^{\prime}) & -h^{\ast}(\tilde{x},\tilde{x}^{\prime}) \end{array} \right]
 \left[\begin{array}{cc} u_{\nu}(\tilde{x}^{\prime}) \\ v_{\nu}(\tilde{x}^{\prime})\end{array} \right]
=E_{\nu} \left[\begin{array}{cc} u_{\nu}(\tilde{x})\\ v_{\nu}(\tilde{x})\end{array} \right], \qquad (\tilde{x}>0)
\end{align}
and $\gamma_{\nu}$ ($\gamma^{\dagger}_{\nu}$) represents the annihilation (creation) operator of a Bogoliubov quasiparticle having the energy $E_{\nu}$.
In what follows, we assume
\begin{align}
0<E_{\nu}<E_{\nu+1}. \qquad (\nu \geq 1)
\end{align}
The BdG Hamiltonian $H_S$ is rewritten as
\begin{align}
\begin{split}
&H_S = \frac{1}{2} [ \boldsymbol{\gamma}^{\dagger}, \boldsymbol{\gamma}^{\mathrm{T}} ] \mathcal{H}
 \left[\begin{array}{cc} \boldsymbol{\gamma} \\ \boldsymbol{\gamma}^{\ast}\end{array} \right],\\
&\boldsymbol{\gamma} = [\gamma_1,\gamma_2,\cdots,\gamma_i,\gamma_{i+1},\cdots]^{\mathrm{T}},\\
&\mathcal{H}=\left[\begin{array}{cc} \mathcal{E} & 0\\ 0 & -\mathcal{E} \end{array} \right],\quad
\mathcal{E}= \mathrm{diag}[E_1,E_2,\cdots,E_i,E_{i+1},\cdots].
\end{split}
\end{align}
The tunneling Hamiltonian $H_T$ is rewritten as
\begin{align}
\begin{split}
&H_T=\frac{1}{2}\left( [ \boldsymbol{\gamma}^{\dagger}, \boldsymbol{\gamma}^{\mathrm{T}} ] \mathcal{W}
 \left[\begin{array}{cc} a_0 \\ a^{\dagger}_0\end{array} \right] + \mathrm{h.c.} \right),\\
&\mathcal{W}=\left[\begin{array}{cc} W \\ \underline{W} \end{array} \right],\quad
W = \left[\begin{array}{cccccc} W_1 \\  W_2 \\ \vdots \\ W_i \\ W_{i+1} \\ \vdots \end{array} \right], \quad
\underline{W} = \left[\begin{array}{cccccc} \underline{W}_1 \\  \underline{W}_2 \\ \vdots \\ \underline{W}_i \\ \underline{W}_{i+1} \\ \vdots \end{array} \right],\\
&W_{\nu} = [-t_Tu^{\ast}_{\nu}(1), t_Tv^{\ast}_{\nu}(1)], \quad
\underline{W}_{\nu} = [-t_Tv_{\nu}(1), t_Tu_{\nu}(1)].
\end{split}
\end{align}
By using $\mathcal{H}$ and $\mathcal{W}$, we can calculate the scattering matrix by~\cite{beenakker_08,flensberg_20}
\begin{align}
S(E) =\left[ \begin{array}{cc}s^{ee}(E) & s^{eh}(E) \\ s^{he}(E) & s^{hh}(E) \end{array} \right]
= 1 - 2 \frac{i}{t} \mathcal{W}^\dagger \left[ E - \mathcal{H} + \frac{i}{t} \mathcal{W}\mathcal{W}^\dagger \right]^{-1} \mathcal{W},
\end{align}
where $s^{ee}$ and $s^{he}$ ($s^{eh}$ and $s^{hh}$) represents the scattering coefficients from the electron (hole) to the electron and hole, respectively.
Here we assume that the eigen states of $\nu=1$ corresponds to the ABS having energy $E_B$ (i.e., $E_1 = E_B$).
In addition, we assume that the ABSs is energetically separated from other states as
\begin{align}
E_2-E_B \gg \frac{\left| t_Tu_{\nu}(1)\right|^2}{t},\quad E_2-E_B \gg \frac{\left| t_Tv_{\nu}(1)\right|^2}{t},
\label{eq:sepe_cnt}
\end{align}
and
\begin{align}
2E_B \gg \frac{\left| t_Tu_{\nu}(1)\right|^2}{t},\quad 2E_B \gg \frac{\left| t_Tv_{\nu}(1)\right|^2}{t},
\label{eq:sepe_abs}
\end{align}
where Eq.~(\ref{eq:sepe_cnt}) characterizes the energy difference between the ABS and the continuum states,
and Eq.~(\ref{eq:sepe_abs}) characterizes the energy difference between the positive-energy and negative-energy ABSs.
These conditions are satisfied in sufficiently low transparency junctions ( i.e., $t_T/t \ll 1$).
For $E \approx E_B$, we can approximately obtain the scattering matrix as~\cite{beenakker_08,feigelman_13,flensberg_20}
\begin{align}
\begin{split}
&S(E) \approx 1 - 2 \frac{i}{t} W_1^\dagger \left[ E - E_B + \frac{i}{t} W_1W_1^\dagger \right]^{-1} W_1
= 1 - \frac{2i}{E-E_B + i (|A_e|^2+|A_h|^2)}\left[ \begin{array}{cc}|A_e|^2 & -A_eA_h^{\ast} \\ -A_e^{\ast}A_h & |A_h|^2 \end{array} \right],\\
&A_e =-\frac{t_Tu_1(1)}{\sqrt{t}} = -\frac{t_Tu^B_1}{\sqrt{t}}, \quad A_h =-\frac{t_Tv_1(1)}{\sqrt{t}}= -\frac{t_Tv^B_1}{\sqrt{t}}.
\end{split}
\end{align}
Within the Blonder--Tinkham--Klapwijk formalism, the differential conductance for bias voltages below the superconducting gap is given by
\begin{align}
G(eV) = \frac{2e^2}{h} \left|s^{he}(E)\right|^2_{E=eV}.
\end{align}
Therefore, for $eV \approx E_B$, we obtain
\begin{align}
G(eV) \approx \frac{2e^2}{h} \frac{4 (|A_e||A_h|)^2}{(eV-E_B)^2+(|A_e|^2+|A_h|^2)^2}= \frac{2e^2}{h}\frac{\Gamma^2}{(eV-E_B)^2 + \Gamma^2} \frac{4 n^e_1 n^h_1}{(n_1)^2},
\label{eq:gns}
\end{align}
where $\Gamma = (t_T^2/t) n_1$.
The equation~(\ref{eq:gns}) corresponds to Eq.~(13) in the main text.

\section{Superconductor UTe$_2$}
\subsection{Bogoliubov--de Gennes Hamiltonian including the spin-orbit coupling potential}
\begin{figure}[t]
\begin{center}
\includegraphics[width=0.325\textwidth]{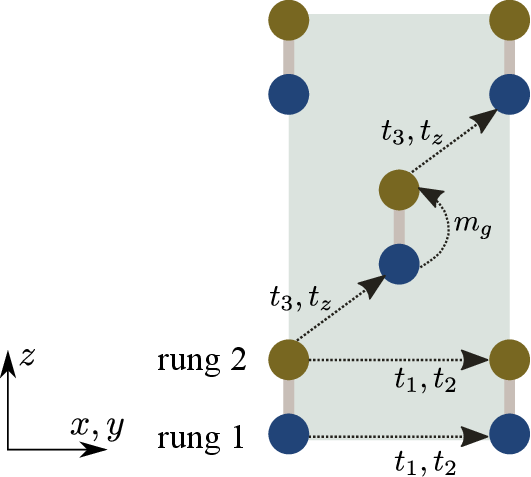}
\caption{\label{fig:sm4}%
Schematic image of the hopping terms in the superconductor UTe$_2$.}
\end{center}
\end{figure}

In this section, we show the BdG Hamiltonian for the superconductor UTe$_2$, which is proposed in Ref.~[\onlinecite{agterberg_21}]:
\begin{align}
\begin{split}
&H_{\mathrm{all}} = H+ H_{\mathrm{soc}},\\
&H = \frac{1}{2}\sum_{\boldsymbol{k},\sigma}[\psi^{\dagger}_{\boldsymbol{k}\sigma},\psi^{\mathrm{T}}_{-\boldsymbol{k}\bar{\sigma}}]\bar{H}_{\boldsymbol{k}}
\left[\begin{array}{cc}\psi_{\boldsymbol{k}\sigma}\\ \psi^{\ast}_{-\boldsymbol{k}\bar{\sigma}}\end{array}\right] ,\quad
\psi^{\dagger}_{\boldsymbol{k}\sigma}=[\psi^{\dagger}_{\boldsymbol{k},\sigma,a,1},\psi^{\dagger}_{\boldsymbol{k},\sigma,a,2},
\psi^{\dagger}_{\boldsymbol{k},\sigma,b,1},\psi^{\dagger}_{\boldsymbol{k},\sigma,b,2}],\\
&\bar{H}_{\boldsymbol{k}}=\left[\begin{array}{cc}\check{h}_{\boldsymbol{k}} & \check{h}_\Delta \\
\check{h}^{\dagger}_\Delta & -\check{h}^{\ast}_{-\boldsymbol{k}} \end{array}\right],\quad
\check{h}_{\boldsymbol{k}}=
\left[\begin{array}{cc} \hat{K}_{\boldsymbol{k}} & \hat{T}^{\dagger}_{\boldsymbol{k}} \\
\hat{T}_{\boldsymbol{k}} & \hat{K}_{\boldsymbol{k}} \end{array} \right],\quad
\check{h}_{\Delta}=\left[\begin{array}{cc} \hat{\Delta} & 0 \\ 0 & \hat{\Delta} \end{array} \right],\\
&\hat{K}_{\boldsymbol{k}}=\left[ \begin{array}{cc} \epsilon_{\boldsymbol{k}} & m_g \\ m_g & \epsilon_{\boldsymbol{k}} \end{array}\right],\quad
\hat{T}_{\boldsymbol{k}}=e^{i\frac{k_x}{2}}e^{i\frac{k_y}{2}}e^{i\frac{k_z}{2}}
\left[ \begin{array}{cc} 0 & f_{g,\boldsymbol{k}}-if_{z,\boldsymbol{k}} \\ f_{g,\boldsymbol{k}}+if_{z,\boldsymbol{k}} & 0 \end{array}\right],\quad
\hat{\Delta} = \left[ \begin{array}{cc} 0 & \Delta \\ -\Delta & 0 \end{array} \right],\\
&\epsilon_{\boldsymbol{k}}=t_1 \cos (k_x) + t_2 \cos (k_y) -\mu,\\
&f_{g,\boldsymbol{k}}=t_3 \cos(k_x/2)\cos(k_y/2)\cos(k_z/2),\\
&f_{z,\boldsymbol{k}}=t_z \cos(k_x/2)\cos(k_y/2)\sin(k_z/2),
\label{eq:ute2_hm0}
\end{split}
\end{align}
where $\psi_{\boldsymbol{k},\sigma,\alpha,l}$ is the annihilation operator of an electron with momentum $\boldsymbol{k}$ and spin $\sigma$ at the $l$-th rung of the sublattice $\alpha$.
The index $\bar{\sigma}$ denotes the opposite spin of $\sigma$.
The each hoping term is shown schematically in Fig.~\ref{fig:sm4}.
According to Ref.~[\onlinecite{agterberg_21}], we choose the fitting parameters as
$(\mu,$ $t_1,$ $t_2,$ $m_0,$ $t_3,$ $t_z)$$=$$(-0.129,$ $-0.0892,$ $0.0678,$ $-0.062,$ $0.0742,$ $-0.0742)$.
We assume the inter-rung-odd spin-triplet $s$-wave pair potential belonging to $A_u$ symmetry, where we choose $\Delta=0.001$.
The spin-orbit coupling potential, which is ignored in the main text, is described by
\begin{align}
\begin{split}
&H_{\mathrm{soc}}=\frac{1}{2}\sum_{\boldsymbol{k}}
[\psi^{\dagger}_{\boldsymbol{k}\uparrow},\psi^{\dagger}_{\boldsymbol{k}\downarrow},
\psi^{\mathrm{T}}_{-\boldsymbol{k}\uparrow},\psi^{\mathrm{T}}_{-\boldsymbol{k}\downarrow}]
\tilde{\Lambda}_{\boldsymbol{k}}
\left[\begin{array}{cc}\psi_{\boldsymbol{k}\uparrow}\\ \psi_{\boldsymbol{k}\downarrow} \\
\psi^{\ast}_{-\boldsymbol{k}\uparrow}\\ \psi^{\ast}_{-\boldsymbol{k}\downarrow}\end{array}\right],\\
&\tilde{\Lambda}_{\boldsymbol{k}}=\left[\begin{array}{cc}\bar{\lambda}_{\boldsymbol{k}} & 0 \\
0 & -\bar{\lambda}^{\ast}_{-\boldsymbol{k}} \end{array}\right],\quad
\bar{\lambda}_{\boldsymbol{k}} =
\left[\begin{array}{cc}\check{\lambda}_{\boldsymbol{k},z} & \check{\lambda}_{\boldsymbol{k},x}-i\check{\lambda}_{\boldsymbol{k},y} \\
\check{\lambda}_{\boldsymbol{k},x}+i\check{\lambda}_{\boldsymbol{k},y} & -\check{\lambda}_{\boldsymbol{k},z} \end{array}\right],\\
&\check{\lambda}_{\boldsymbol{k},x(y)}=\left[\begin{array}{cc}\hat{\lambda}_{\boldsymbol{k},x(y)} & 0 \\ 0 & \hat{\lambda}_{\boldsymbol{k},x(y)} \end{array}\right], \quad
\hat{\lambda}_{\boldsymbol{k},x(y)}=
\left[\begin{array}{cc}t_{x(y)} \sin(k_{x(y)}) & 0 \\ 0 & -t_{x(y)} \sin(k_{x(y)}) \end{array}\right],\\
&\check{\lambda}_{\boldsymbol{k},z}=\left[\begin{array}{cc} 0 & \hat{\lambda}^{\dagger}_{\boldsymbol{k},z} \\ \hat{\lambda}_{\boldsymbol{k},z} & 0 \end{array}\right], \quad
\hat{\lambda}_{\boldsymbol{k},z}=e^{i\frac{k_x}{2}}e^{i\frac{k_y}{2}}e^{i\frac{k_z}{2}}
\left[\begin{array}{cc}f_{u,\boldsymbol{k}}& 0 \\ 0 & -f_{u,\boldsymbol{k}} \end{array}\right],\\
& f_{u,\boldsymbol{k}}=t_u \sin(k_x/2)\sin(k_y/2)\sin(k_z/2),
\end{split}
\end{align}
where $(t_x,$ $t_y,$ $t_u)$$=$$(0.006,$ $0.008,$ $0.01)$~\cite{agterberg_21}.

\begin{figure}[h]
\begin{center}
\includegraphics[width=0.5\textwidth]{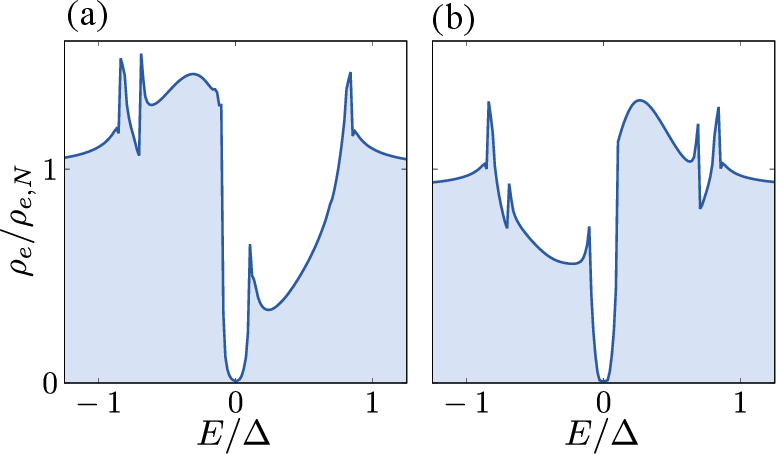}
\caption{\label{fig:sm5}%
Electron part of the spectral function in the presence of the spin-orbit coupling potential as a function of the energy $E$.
The pair potential belongs to $A_u$ pairing symmetry.
In (a) and (b), we show the result for the outermost site, i.e., $\tilde{y}=1$, and that for the second outermost site, i.e., $\tilde{y}=1+\frac{1}{2}$, respectively.
}
\end{center}
\end{figure}

Here, we discuss the electron part of the spectral function in the presence of the spin-orbit coupling potential:
\begin{align}
&\rho_{e}(\tilde{y},E)=\sum_{k_x,k_z} \rho_{e}(\tilde{y},k_x,k_z,E),\quad
\rho_e(\tilde{y},k_x,k_z,E)=-\frac{1}{\pi}\mathrm{Im}[g_{k_x,k_z}(\tilde{y},\tilde{y},E)],
\end{align}
where $\tilde{y}=y$ or $y+\frac{1}{2}$ measures the distance from the surface with $y$ being an integer,
and $g_{k_x,k_z}(\tilde{y},\tilde{y},E)$ is the retarded Green's function obtained using the recursive Green's function techniques~\cite{fisher_81,ando_91}.
In Figs.~\ref{fig:sm5}(a), we show $\rho_{e}(\tilde{y},E)$ at the outermost site ($\tilde{y}=1$) and the second outermost site ($\tilde{y}=1+\frac{1}{2})$ as a function of the energy, respectively.
We find that $\rho_{e}(\tilde{y},E)$ at the outermost (second-outermost) has the significant hump only for $E<0$ ($E>0$).
Consequently, we confirm the appearance of the OCABSs even with the spin-orbit coupling potential.

\subsection{Band basis Hamiltonian at $k_z=\pi$}
In this section, we calculate the band basis Hamiltonian at $k_z=\pi$, which is given by Eq.~(16) in the main text.
For this calculation, we ignore the spin-orbit coupling term $H_{\mathrm{soc}}$.
At $k_z=\pi$, the Hamiltonian $\bar{H}_{\boldsymbol{k}}$ in Eq.~(\ref{eq:ute2_hm0}) is rewritten as
\begin{align}
\begin{split}
&\bar{H}_{\boldsymbol{q}}=\left[\begin{array}{cc}\check{h}_{\boldsymbol{q}} & \check{h}_\Delta \\
\check{h}^{\dagger}_\Delta & -\check{h}^{\ast}_{-\boldsymbol{q}} \end{array}\right], \\
&\check{h}_{\boldsymbol{q}}=\left[\begin{array}{cccc}
\epsilon_{\boldsymbol{q}} & m_g & 0 & - \chi^{\ast}_{\boldsymbol{q}} f_{z,\boldsymbol{q}}\\
m_g & \epsilon_{\boldsymbol{q}} & \chi^{\ast}_{\boldsymbol{q}} f_{z,\boldsymbol{q}} & 0 \\
0 & \chi_{\boldsymbol{q}} f_{z,\boldsymbol{q}} & \epsilon_{\boldsymbol{q}} & m_g \\
-\chi_{\boldsymbol{q}} f_{z,\boldsymbol{q}} & 0 & m_g & \epsilon_{\boldsymbol{q}} \end{array}\right],\quad
\chi_{\boldsymbol{q}} = e^{i\frac{k_x}{2}}e^{i\frac{k_y}{2}},
\end{split}
\end{align}
where $\boldsymbol{q}=(k_x,k_y,k_z=\pi)$.
By using a unitary operator,
\begin{align}
\begin{split}
&\bar{U} = \bar{U}_4 \bar{U}_3 \bar{U}_2 \bar{U}_1, \\
&\bar{U}_1 = \left[\begin{array}{cc}\check{U}_1 & 0 \\ 0 & \check{U}_1 \end{array}\right],\quad
\check{U}_1=\frac{1}{\sqrt{2}}\left[\begin{array}{cccc}1&1&0&0\\-1&1&0&0\\0&0&1&1\\0&0&-1&1\end{array}\right],\\
&\bar{U}_2 = \left[\begin{array}{cc}\check{U}_2 & 0 \\ 0 & \check{U}_2 \end{array}\right],\quad
\check{U}_2=\left[\begin{array}{cccc}1&0&0&0\\0&0&0&1\\0&1&0&0\\0&0&1&0\end{array}\right],\\
&\bar{U}_3 = \left[\begin{array}{cccc} \hat{1}&0&0&0\\0&0&0&\hat{1}\\0&\hat{1}&0&0\\0&0&\hat{1}&0\end{array}\right],\quad
\hat{1}=\left[\begin{array}{cc}1 & 0 \\ 0 & 1 \end{array}\right],\\
&\bar{U}_4 = \left[\begin{array}{cc}\check{U}_4 & 0 \\ 0 & \check{U}_4 \end{array}\right],\quad
\check{U}_4=\left[\begin{array}{cccc}1&0&0&0\\0&1&0&0\\0&0&0&1\\0&0&1&0\end{array}\right],
\end{split}
\end{align}
we deform the Hamiltonian $\bar{H}_{\boldsymbol{q}}$ as
\begin{align}
\begin{split}
&\bar{H}^{\prime}_{\boldsymbol{q}}=\bar{U} \bar{H}_{\boldsymbol{q}}\bar{U}^{\dagger} =
\left[\begin{array}{cc}\check{H}^{\prime}_{\boldsymbol{q},+} & 0 \\ 0 & \check{H}^{\prime}_{\boldsymbol{q},-} \end{array}\right],\\
&\check{H}^{\prime}_{\boldsymbol{q},s}=\left[\begin{array}{cccc}
\epsilon_{\boldsymbol{q}}+s m_g & -s \chi^{\ast}_{\boldsymbol{q}}f_{z,\boldsymbol{q}} & 0 & s \Delta \\
-s \chi_{\boldsymbol{q}}f_{z,\boldsymbol{q}} & \epsilon_{\boldsymbol{q}}-s m_g & -s \Delta & 0\\
0 & - s \Delta & -\epsilon_{\boldsymbol{q}}-s m_g & -s \chi_{\boldsymbol{q}}f_{z,\boldsymbol{q}}\\
s \Delta & 0 & -s \chi^{\ast}_{\boldsymbol{q}}f_{z,\boldsymbol{q}} & -\epsilon_{\boldsymbol{q}}+s m_g \end{array}\right].
\end{split}
\end{align}
Furthermore, by using a unitary operator,
\begin{align}
\begin{split}
&\bar{U}_5 = \left[\begin{array}{cc}\check{U}_5 & 0 \\ 0 & \check{U}_5 \end{array}\right], \quad
\check{U}_5 = \left[\begin{array}{cc}\hat{U}_5 & 0 \\ 0 & \underline{\hat{U}}_5 \end{array}\right],\\
&\hat{U}_5 = \frac{1}{\sqrt{2V_{\boldsymbol{q}} (V_{\boldsymbol{q}}+m_g) }}
\left[\begin{array}{cc}V_{\boldsymbol{q}}+m_g & - \chi^{\ast}_{\boldsymbol{q}}f_{z,\boldsymbol{q}} \\
- \chi_{\boldsymbol{q}}f_{z,\boldsymbol{q}} & -(V_{\boldsymbol{q}}+m_g) \end{array}\right],\\
&\underline{\hat{U}}_5 = \frac{1}{\sqrt{2V_{\boldsymbol{q}} (V_{\boldsymbol{q}}+m_g) }}
\left[\begin{array}{cc}V_{\boldsymbol{q}}+m_g & \chi_{\boldsymbol{q}}f_{z,\boldsymbol{q}} \\
\chi_{\boldsymbol{q}}^{\ast} f_{z,\boldsymbol{q}} & -(V_{\boldsymbol{q}}+m_g) \end{array}\right],\\
&V_{\boldsymbol{q}}=\sqrt{m^2_g+f^2_{z,\boldsymbol{q}}},
\end{split}
\end{align}
we eventually obtain the band-basis Hamiltonian,
\begin{align}
\begin{split}
&\bar{H}^{\mathrm{band}}_{\boldsymbol{q}}=\bar{U}_5 \bar{H}^{\prime}_{\boldsymbol{q}}\bar{U}^{\dagger}_5=
\left[\begin{array}{cc} \check{H}_{\boldsymbol{q},+} & 0 \\ 0 & \check{H}_{\boldsymbol{q},-} \end{array}\right]\\
&\check{H}_{\boldsymbol{q},s}=\left[ \begin{array}{cccc}
\epsilon_{\boldsymbol{q}}+sV_{\boldsymbol{q}} & 0 & s\Delta \alpha_{\boldsymbol{q},-} & -s\Delta \beta_{\boldsymbol{q}}\\
0 & \epsilon_{\boldsymbol{q}}-sV_{\boldsymbol{q}} & s\Delta \beta_{\boldsymbol{q}} & s\Delta \alpha_{\boldsymbol{q},+}\\
s\Delta \alpha_{\boldsymbol{q},+} & s\Delta \beta_{\boldsymbol{q}} & -\epsilon_{\boldsymbol{q}}-sV_{\boldsymbol{q}} & 0\\
-s\Delta \beta_{\boldsymbol{q}} & s\Delta \alpha_{\boldsymbol{q},-} & 0 & -\epsilon_{\boldsymbol{q}}+sV_{\boldsymbol{q}}\\
\end{array}\right],\\
&\alpha_{\boldsymbol{q},\pm}=e^{\pm ik_x/2}e^{\pm ik_y/2}\frac{f_{z,\boldsymbol{q}}}{V_{\boldsymbol{q}}},\quad
\beta_{\boldsymbol{q}}=\frac{m_g}{V_{\boldsymbol{q}}},
\end{split}
\end{align}
which corresponds to Eq.~(16) in the main text.

\subsection{Electron part of spectral function with other pairing symmetries}
\begin{figure}[h]
\begin{center}
\includegraphics[width=0.5\textwidth]{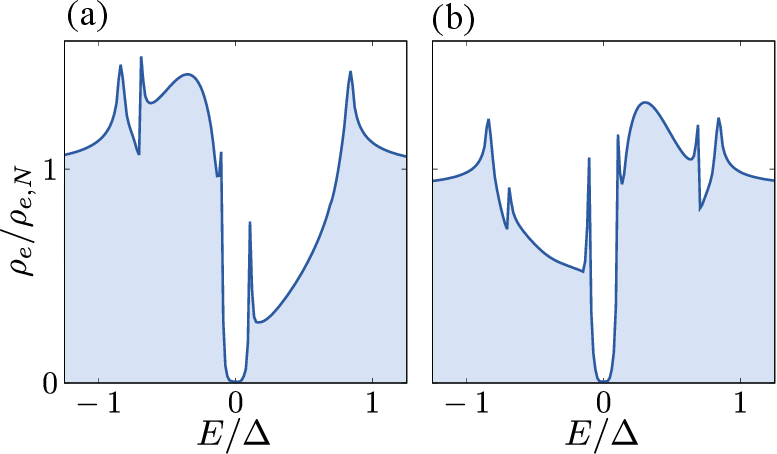}
\caption{\label{fig:sm6}%
Electron part of the spectral function with $B_{2u}$ pairing symmetry as a function of the energy $E$.
In (a) and (b), we show the result for the outermost site, i.e., $\tilde{y}=1$, and that for the second outermost site, i.e., $\tilde{y}=1+\frac{1}{2}$, respectively.
}
\end{center}
\end{figure}

\begin{figure}[h]
\begin{center}
\includegraphics[width=0.5\textwidth]{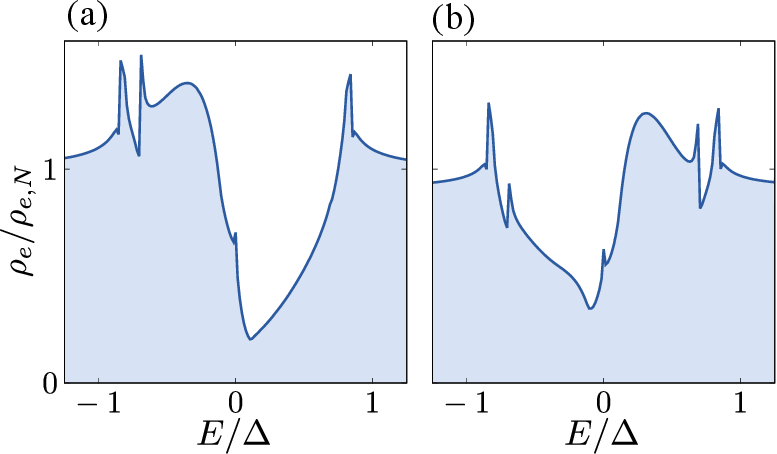}
\caption{\label{fig:sm7}%
Electron part of the spectral function with $B_{3u}$ pairing symmetry as a function of the energy $E$.
In (a) and (b), we show the result for the outermost site, i.e., $\tilde{y}=1$, and that for the second outermost site, i.e., $\tilde{y}=1+\frac{1}{2}$, respectively.
}
\end{center}
\end{figure}

In this section, we show the electron part of spectral function with other possible pairing symmetries discussed in Ref.~[\onlinecite{agterberg_21}].
Specifically, we consider the inter-rung-odd spin-triplet $s$-wave pair potential belonging to $B_{2u}$ and $B_{3u}$ symmetries.
The BdG Hamiltonian is given by
\begin{align}
\begin{split}
&H = \frac{1}{2}\sum_{\boldsymbol{k},\sigma}[\psi^{\dagger}_{\boldsymbol{k}\sigma},\psi^{\mathrm{T}}_{-\boldsymbol{k}\sigma}]\bar{H}^{2u(3u)}_{\boldsymbol{k}}
\left[\begin{array}{cc}\psi_{\boldsymbol{k}\sigma}\\ \psi^{\ast}_{-\boldsymbol{k}\sigma}\end{array}\right] + H_{\mathrm{soc}},\\
&\bar{H}^{2u(3u)}_{\boldsymbol{k}}=\left[\begin{array}{cc}\check{h}_{\boldsymbol{k}} & \check{h}^{2u(3u)}_{\Delta,\sigma} \\
\left\{ \check{h}^{2u(3u)}_{\Delta,\sigma}\right\}^{\dagger} & -\check{h}^{\ast}_{-\boldsymbol{k}} \end{array}\right],\\
&\check{h}^{2u}_{\Delta,\sigma}=-s_{\sigma}\left[\begin{array}{cc} \hat{\Delta} & 0 \\ 0 & \hat{\Delta} \end{array} \right], \quad
\check{h}^{3u}_{\Delta,\sigma}=i\left[\begin{array}{cc} \hat{\Delta} & 0 \\ 0 & \hat{\Delta} \end{array} \right],
\end{split}
\end{align}
where $s_{\uparrow(\downarrow)}=1$ ($-1$).
In the following, we focus on the ABSs at the surface parallel to the $x$ direction.
In Figs.~\ref{fig:sm6}(a) and \ref{fig:sm6}(b), we show $\rho_{e}(\tilde{y},E)$ with $B_{2u}$ pairing symmetry 
at the outermost site ($\tilde{y}=1$) and the second outermost site ($\tilde{y}=1+\frac{1}{2})$ as a function of the energy, respectively.
In Figs.~\ref{fig:sm7}(a) and \ref{fig:sm7}(b), we show $\rho_{e}(\tilde{y},E)$ with $B_{3u}$ pairing symmetry at $\tilde{y}=1$ and $\tilde{y}=1+\frac{1}{2}$ as a function of $E$, respectively.
In both cases, we find that $\rho_{e}(\tilde{y},E)$ at the outermost (second outermost) has the significant hump only for $E<0$ ($E>0$).
As a result, we confirm the appearance of the OCABSs even with $B_{2u}$ and $B_{3u}$ symmetries.

\end{document}